\documentclass[prb,showpacs,twocolumn,aps]{revtex4}
\usepackage[dvipdfm]{graphicx}
\usepackage[dvipdfm]{color}
\usepackage{dcolumn}
\usepackage{amsmath}
\usepackage{amssymb}

\newcommand{\lcco}{$\rm La_{2}CaCu_2O_{6+{\it \delta}}$}
\newcommand{\lsrco}{$\rm La_{2}SrCu_2O_{6+{\it \delta}}$}
\newcommand{\lsrcco}{$\rm La_{2-{\it x}}Sr_{\it x}CaCu_2O_{6+{\it \delta}}$}
\newcommand{\lccco}{$\rm La_{1.9}Ca_{1.1}Cu_2O_{6+{\it \delta}}$}
\newcommand{\lscco}{$\rm La_{1.85}Sr_{0.15}CaCu_2O_{6+{\it \delta}}$}
\newcommand{\lsccco}{$\rm La_{2-{\it x}}(Sr,Ca)_{\it x}CaCu_2O_{6+{\it \delta}}$}
\newcommand{\shibata}{$\rm La_{1.89}Ca_{1.11}Cu_2O_{6+{\it \delta}}$}
\newcommand{\ulrich}{$\rm La_{1.80}Sr_{0.20}CaCu_2O_{6+{\it \delta}}$}

\newcommand{\lco}{$\rm La_2CuO_4$}

\newcommand{\lsco}{$\rm La_{2-{\it x}}Sr_{\it x}CuO_4$}

\newcommand{\scoc}{$\rm Sr_2CuO_2Cl_2$}

\newcommand{\ybco}{$\rm YBa_2Cu_3O_{6+{\it \delta}}$}

\newcommand{\ybccos}{$\rm Y_{1-{\it x}}Ca_{\it x}Ba_2Cu_3O_{6}$}

\newcommand{\ybcot}{$\rm YBa_2Cu_3O_{6.3}$}

\newcommand{\sus}{susceptibility}

\newcommand{\DM}{Dzyaloshinsky--Moriya}

\newcommand{\ohoht}{($\frac{1}{2},\frac{1}{2},3$)}

\include{version}

\begin{document}

\title{Neutron scattering study on $\bf La_{1.9}Ca_{1.1}Cu_2O_{6+{\it \delta}}$ and $\bf La_{1.85}Sr_{0.15}CaCu_2O_{6+{\it \delta}}$}

\author{M. H\"ucker}
\author{Young-June Kim}
\author{G. D. Gu}
\affiliation{Physics Department, Brookhaven National Laboratory, Upton, New York 11973, USA}
\author{B. D.\ Gaulin}
\affiliation{Department of Physics and Astronomy, McMaster University, Hamilton, Ontario, Canada L8S 4M1}
\author{J. W.\ Lynn}
\affiliation{NIST Center for Neutron Research, National Institute of Standards and Technology, Gaithersburg, Maryland 20742, USA}
\author{J. M. Tranquada}
\affiliation{Physics Department, Brookhaven National Laboratory, Upton, New York 11973, USA}

\date{\today}

\begin{abstract}

We present neutron scattering data on two single crystals of the high
temperature superconductor $\rm La_{2-{\it x}}(Ca,Sr)_{\it
x}CaCu_2O_{6+{\it \delta}}$. The $\rm Ca_{0.1}$-doped crystal
exhibits a long-range antiferromagnetically ordered ground state. In
contrast, the $\rm Sr_{0.15}$-doped crystal exhibits short-range
antiferromagnetic order as well as weak superconductivity. In both
crystals antiferromagnetic correlations are commensurate; however,
some results on the $\rm Ca_{0.1}$-doped crystal resemble those on
the spin-glass phase of \lsco , where magnetic correlations became
incommensurate. In addition, both crystals show a structural
transition from tetragonal to orthorhombic symmetry. Quite
remarkably, the temperature dependence and correlation length of the
magnetic order is very similar to that of the orthorhombic
distortion. We attribute this behavior to an orthorhombic
strain-induced inter-bilayer magnetic coupling, which triggers the
antiferromagnetic order. The large size of the crystals made it also
possible to study the magnetic diffuse scattering along rods
perpendicular to the $\rm CuO_2$ planes in more detail. For
comparison we show X-ray diffraction and magnetization data. In
particular, for the $\rm Ca_{0.1}$-doped crystal these measurements
reveal valuable information on the spin-glass transition as well as a
second anomaly associated with the N\'eel transition.

\end{abstract}


\maketitle

\section{Introduction}
\label{intro}
The system \lcco\ is a very interesting member of the family of high
temperature superconductors. Similar to \ybco\ it has $\rm CuO_2$
bilayers. However, it has no $\rm CuO$ chain-layers.~\cite{Cava90b}
To introduce hole-like charge carriers into the $\rm CuO_2$ planes it
can be doped with Sr and Ca, although specimens generally have to be
annealed under high oxygen pressure to become bulk
superconductors.~\cite{Cava90c,Kinoshita92b,Klamut00a} The maximum
$T_c$ of $\sim$60~K is relatively low compared to other bilayer
systems.~\cite{Pavarini01a} High pressure oxygen annealing was shown
to introduce interstitial oxygen, which leads to an increase of the
hole concentration.~\cite{Kinoshita92b,Shaked93a} Furthermore, it
increases the miscibility range for Sr and Ca
doping.~\cite{Kinoshita92b} The detailed role of the oxygen
interstitials, in particular with respect to superconductivity, is
still unclear.

So far, most studies were performed on polycrystalline specimens,
which are much easier to homogenously charge with oxygen than are the
large single crystals needed for neutron scattering. However, the
study of underdoped crystals grown at low oxygen pressure is as
important as the study of bulk superconductors, since high
temperature superconductivity at high hole doping seems to be
intimately connected with the antiferromagnetic correlations at low
hole doping.~\cite{Orenstein00a} In particular, there is growing
evidence that incommensurate antiferromagnetic correlations of the Cu
spins are associated with the so-called charge and spin stripes,
which may play a vital role for
superconductivity.~\cite{Cheong91,Mason92,Hayden92aN,
Tranquada95a,Mook02a,Dai01a,Zaanen00a,Kivelson03a,Tranquada04a,Hayden04a}
In a recent neutron diffraction experiment on a non-superconducting
\ulrich\ single crystal grown at 1~atm oxygen pressure a commensurate
short-range antiferromagnetic order was observed.~\cite{Ulrich02a}
This finding is in sharp contrast to the incommensurate spin stripes
found in \lsco\ with comparable hole
concentration.~\cite{Wakimoto00a,Yamada98a} The present study is
aimed at a better understanding of the different behavior in \lsccco
.

\begin{figure}[t]
\center{\includegraphics[width=0.45\columnwidth,angle=0,clip]{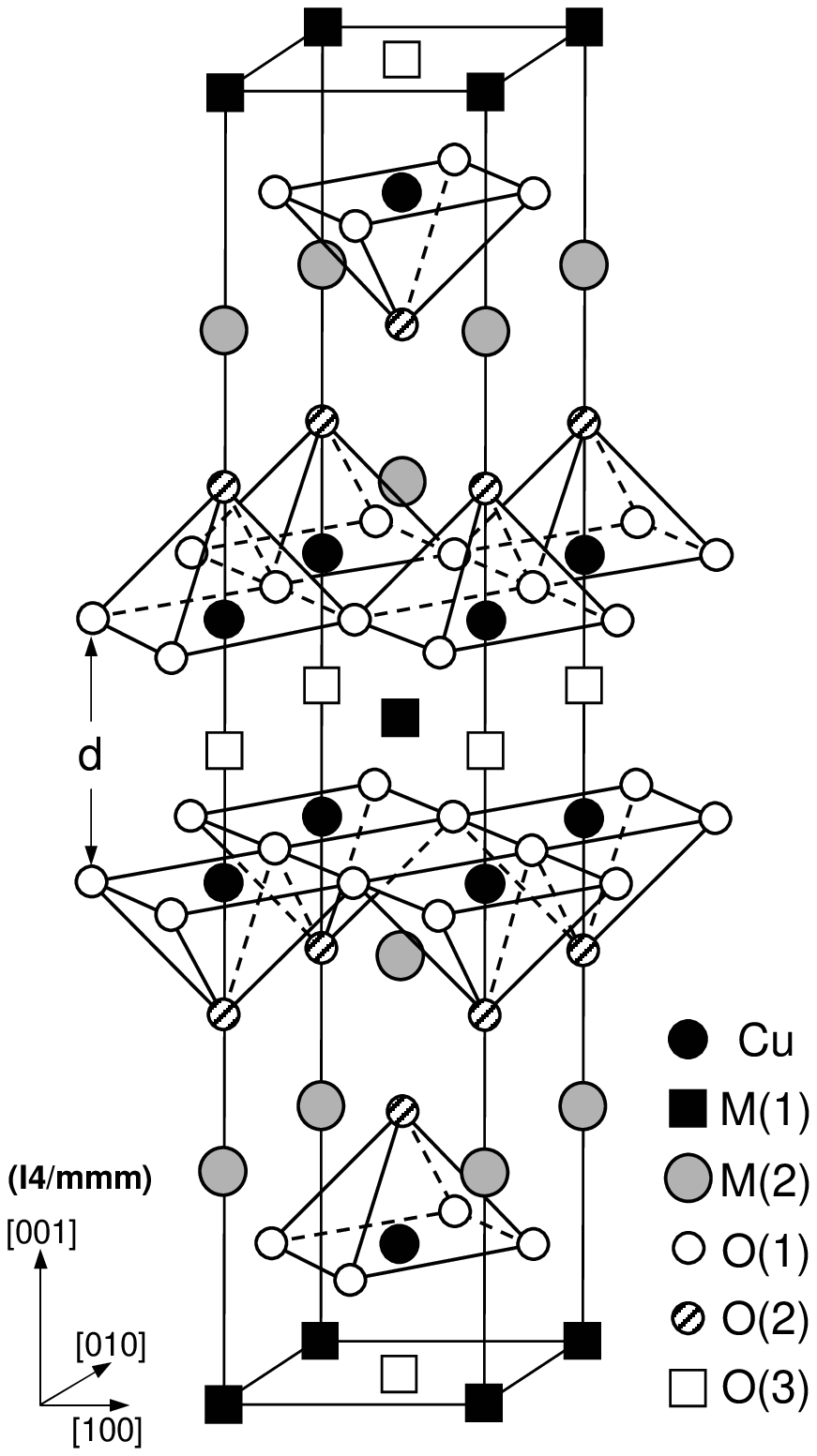}}
\caption[]{Unit cell of \lsccco\ in the tetragonal high temperature
phase.} \label{fig1}
\end{figure}

Structurally, \lsccco\ is known as the most simple bilayer
system.~\cite{Cava90b,Cava90c} Its unit cell, displayed in
Fig.~\ref{fig1}, can be derived from that of the single layer sister
compound \lsco\ by replacing the $\rm CuO_6$ octahedral network by
two $\rm CuO_5$ pyramidal planes, separated by a cation monolayer. In
a stoichiometric compound ($\delta= 0$) the interstitial oxygen site
O(3) is not occupied.~\cite{oxygen} The structure of the rocksalt
layers separating adjacent bilayers is identical to those in \lsco .
In the undoped case ($x=0$), one would ideally expect that the
lattice site M(1) in the center of a bilayer is occupied by Ca, and
the lattice site M(2) in the rock salt layers by La
[Fig.~\ref{fig1}]. However, it turns out that about 10\% of the M(1)
sites are occupied by La, while a corresponding number of alkaline
earths go on the M(2)
site.~\cite{Shaked93a,Ohyama95a,Deng99b,Ulrich02a} This ratio will of
course change slightly upon doping Sr or Ca for La. Interstitial
oxygen introduced by high pressure oxygen annealing goes on the O(3)
site, and effectively bridges two $\rm CuO_5$ pyramids in a bilayer
to two $\rm CuO_6$ octahedra, coupled via their apical
oxygen.~\cite{Izumi89a,Grasmeder90a,Lightfoot90a,
Kinoshita92b,Shaked93a}

In this work we present neutron scattering results on two large
single crystals with the compositions \lccco\ and \lscco . Optical
conductivity measurements by Wang et al.~\cite{Wang03a} have shown
that both compounds contain a significant hole concentration $p$ per
$\rm CuO_2$ plane, roughly consistent with the nominal values
$p=x/2$. [Note that for \lsco\ $p=x$.] From neutron diffraction (ND)
we find that the Ca-doped crystal exhibits a long-range
antiferromagnetic order. In contrast, the Sr-doped crystal exhibits a
short-range antiferromagnetic order as well as weak
superconductivity, with $T_{\rm c}^{\rm onset}\sim 25$~K and a
superconducting volume fraction of the order of one percent. In both
crystals antiferromagnetism is commensurate, which is not surprising
for the long-range order, but somewhat unexpected in the case of the
Sr-doped crystal. In addition, both crystals show a structural
transition from a high-temperature-tetragonal phase to a
low-temperature-orthorhombic phase, first identified in
Ref.~\onlinecite{Ulrich02a}. The temperature dependence of the
magnetic order and the orthorhombic lattice distortion show a
remarkable similarity, which we discuss in terms of a distortion
induced magnetic inter-bilayer coupling. It is possible that this
magneto-elastic coupling influences the electronic ground state and
maybe even determines whether it is magnetic or superconducting. The
large size of the crystals made it possible to study the elastic and
inelastic response from two dimensional (2D) diffuse scattering,
which shows up in rods along $\it Q_z$. Furthermore, we have
performed a number of X-ray diffraction (XRD) measurements, which
reveal additional information about the temperature dependence of the
line width of the orthorhombic superlattice peak in both crystals
(Sec.~\ref{xrays}). Measurements of the static magnetic \sus\ reveal
information about the spin-glass transition, and show a second
anomaly associated with the antiferromagnetic order in the Ca-doped
crystal (Sec.~\ref{magnetism}).

\section{Experimental}
\label{exp}
The two centimeter-size single crystals with $\varnothing$ 6-7~mm and
length of $\sim$10~cm were grown at Brookhaven and McMaster University by the
travelling-solvent floating-zone method. The \lccco\ crystal was
grown in an atmosphere of flowing oxygen gas at a pressure of $p({\rm
O_2})=1$~atm. The \lscco\ crystal was grown at $p({\rm O_2})=11$~atm,
to increase the concentration of oxygen interstitials. At the given
pressures, both compositions are within the range for single-phase
solid solutions of this material.~\cite{Kinoshita92b} Our attempts to
grow crystals with $x \geq 0.2$ and $p({\rm O_2})=11$~atm inevitably
yielded samples containing a second phase.

In the case of \lccco , neutron scattering experiments were performed
on two large pieces with weights of 3.9~g and 5.3~g. For \lscco\ the
studied piece had a weight of 5.1~g. As mentioned earlier, this
crystal shows weak superconductivity, although it has a lower Sr
content than the crystal studied in Ref.~\onlinecite{Ulrich02a},
which was grown at $p({\rm O_2})=1$~atm and insulating at low
temperatures. We associate this different behavior with the higher
oxygen pressure used in our crystal growth procedure.

\begin{figure}[t]
\center{\includegraphics[width=0.90\columnwidth,angle=0,clip]{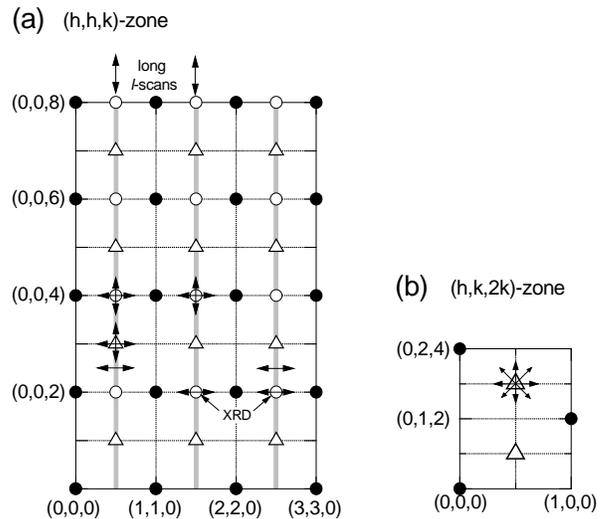}}
\caption[]{Reciprocal space for the $(h,h,l)$ and $(h,k,2k)$ zones
with typical scans indicated by arrows. X-ray diffraction experiments
(XRD) where mainly performed at $(\frac{3}{2},\frac{3}{2},2)$ and
$(\frac{5}{2},\frac{5}{2},2)$.} \label{fig2}
\end{figure}

The neutron scattering experiments were carried out on the
triple-axis spectrometer BT9 at the NIST Center for Neutron Research,
at neutron energies of 14.7~meV and 30.5~meV. Pyrolytic graphite (PG)
(002) reflections were used for both the monochromator and the
analyzer. To eliminate higher-order contamination, an additional PG
filter was put into the beam after the analyzer. The crystals were
mounted in a He filled Al can. X-ray diffraction experiments were
performed in reflection at beamline X22C of the National Synchrotron
Light Source at Brookhaven at a photon energy of 8.9~keV. The
magnetization of small pieces of the crystals was measured with a
SQUID (superconducting quantum interference device) magnetometer.

\section{Results and Discussion}
\label{results}
Figure~\ref{fig2}(a) shows the reciprocal space for the $(h,h,l)$
zone of \lcco\ in the notation of the tetragonal unit cell of
Fig.~\ref{fig1} with space group $I4/mmm$. At room temperature both
crystals are tetragonal and only fundamental Bragg peaks ($\bullet$)
are observed. At low temperatures additional peaks appear, indicating
the transition into the orthorhombic phase~\cite{Ulrich02a} ($\circ$)
as well as static antiferromagnetic order ($\vartriangle$). Note that
magnetic Bragg peaks are allowed for $l=n$ ($n\neq 0$) but for $l$
even they are dominated by the nuclear superlattice
peaks.~\cite{Freltoft87,Tranquada88b} In the orthorhombic phase, with
space group $Bmab$, no splitting of the in-plane lattice constants
was observed within the resolution of the neutron and X-ray
experiments, although X-ray data show a clear broadening of
corresponding fundamental Bragg peaks for in-plane momentum
transfers. This suggests that the orthorhombic strain is weak, i.e.,
much weaker than in \lsco .~\cite{Boeni88,Radaelli94} Therefore,
throughout this paper reflections will be indexed using the notation
for the tetragonal unit cell and scattering vectors ${\bf Q}=(h,k,l)$
will be specified in units of $(2\pi/a, 2\pi/a, 2\pi/c)$. The thick
grey lines in Fig.~\ref{fig2}(a) symbolize scattering rods along the
$c$-axis from 2D scattering. The arrows mark the positions where most
of the scans were performed. To test whether the magnetic Bragg peaks
are incommensurate, the crystal was mounted in the $(h,k,2k)$ zone
where all scans were performed on the ($\frac{1}{2},\frac{3}{2},3)$
peak [see Fig.~\ref{fig2}(b)]. For the lattice parameters at
$\sim$10~K, as determined by neutron diffraction, we find $a =
3.82(1)$~\AA\ and $c=19.41(5)$~\AA\ for \lscco , as well as $a =
3.83(1)$~\AA\, and $c=19.36(5)$~\AA\ for \lccco .

\begin{figure}[t]
\center{\includegraphics[width=1.1\columnwidth,angle=270,clip]{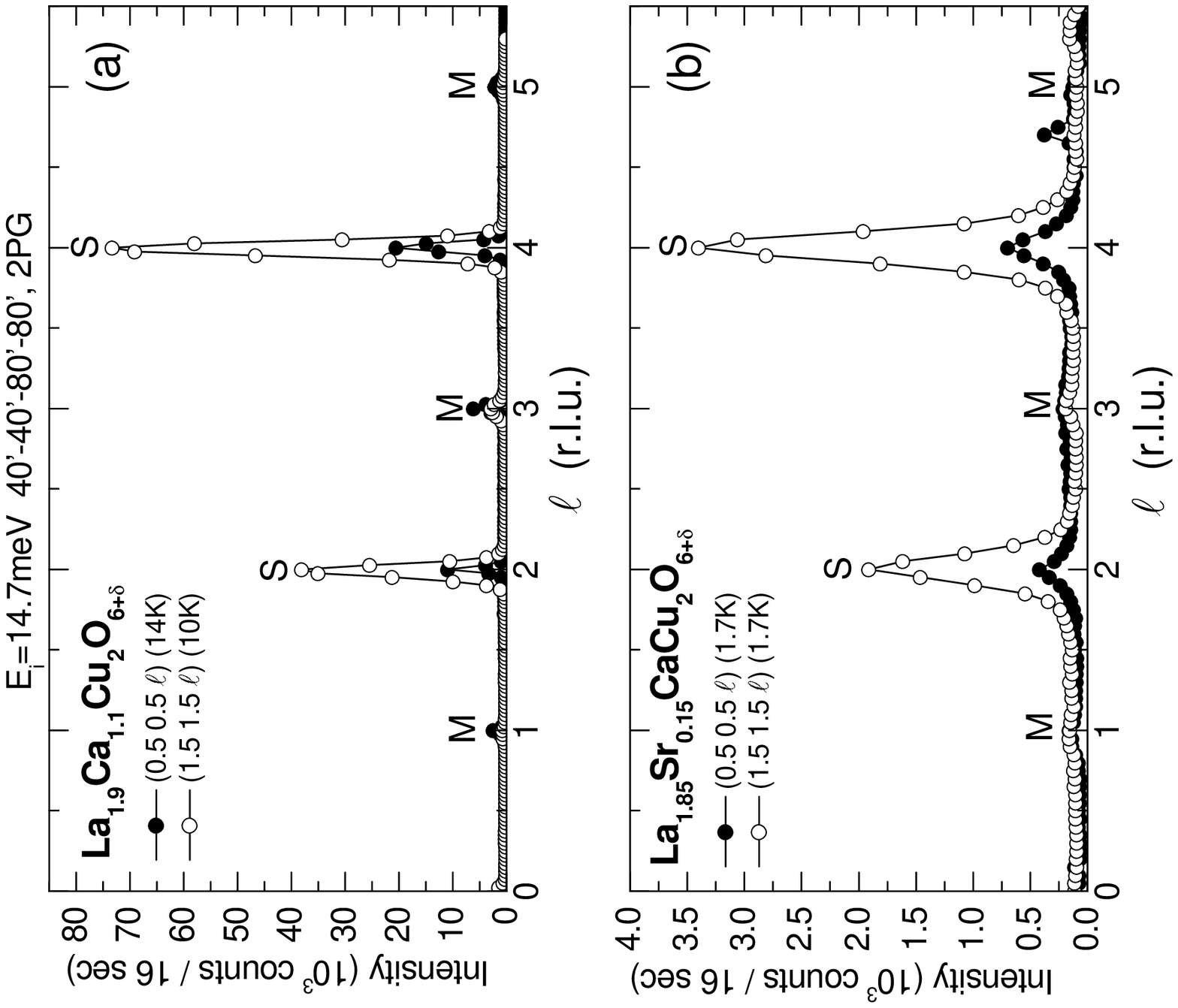}}

\caption[]{Elastic scans along ${\bf Q} =
(\frac{1}{2},\frac{1}{2},l)$ and $(\frac{3}{2},\frac{3}{2},l)$ for
\lccco\ (top) and \lscco\ (bottom). M indicates magnetic Bragg
reflections and S predominantly nuclear superlattice reflections.}
\label{fig3}
\end{figure}

\subsection{Neutron Scattering}
\subsubsection{Magnetic Bragg and orthorhombic superlattice peaks}
\label{elastic}
In Fig.~\ref{fig3}(a) we show long {\it l}-scans along ${\bf Q} =
(\frac{1}{2},\frac{1}{2},l)$ and $(\frac{3}{2},\frac{3}{2},l)$ for
\lccco\ at low temperatures [cf. Fig.~\ref{fig2}(a)]. One can clearly
see the increase of the predominantly nuclear superlattice peaks for
$l=2n$ ($n\neq 0$) as well as the decrease of the purely magnetic
Bragg peaks for $l$ odd with increasing $|{\bf Q}|$. As mentioned
earlier there is also a small magnetic contribution for $l$ even.
Weak intensity observed at $l=0$ is believed to arise from higher
harmonics or multiple scattering. Note that in contrast to \lco\ no
magnetic peak is expected at $l=0$, since the magnetic structure
factor is zero (see
Sec.~\ref{diffuse}).~\cite{Freltoft87,Tranquada88b} The peak width in
Fig.~\ref{fig3}(a) is resolution limited. From $h$-scans through the
magnetic $(\frac{1}{2},\frac{1}{2},1)$ peak with a collimation of
$10'$-$10'$-$10'$-$10'$ (not shown) we have extracted a minimum
in-plane correlation length of $\rm \xi \simeq 270$~\AA , suggesting
that \lccco\ exhibits a long-range antiferromagnetic
order.~\cite{correl} In the case of \lscco\ in Fig.~\ref{fig3}(b) the
peak intensities are about one order of magnitude smaller and the
peak widths much larger, indicating a short-range order. For the
structural distortion we find that in-plane and out-of-plane
correlation lengths are about the same and of the order of $\rm \xi
\simeq 27$~\AA. Interestingly, the in-plane correlation length of the
magnetic order is nearly identical with this, as will be shown below
in more detail. However, the magnetic peaks are too weak to extract a
reliable value for the correlation length along $c$.

\begin{figure}[t]
\center{\includegraphics[width=1\columnwidth,angle=0,clip]{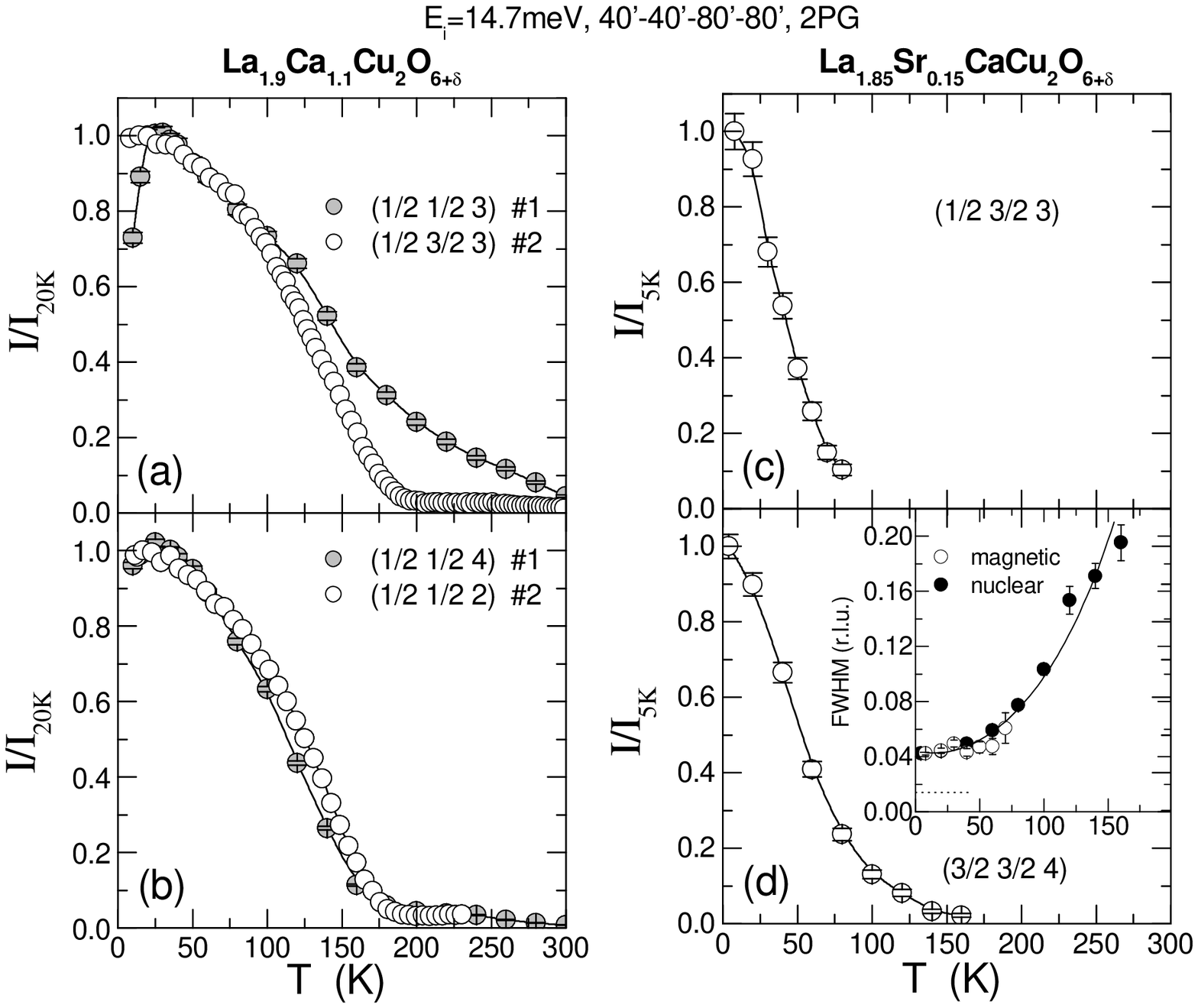}}
\caption[]{Intensity of the magnetic Bragg peaks
$(\frac{1}{2},\frac{1}{2},3)$ and $(\frac{1}{2},\frac{3}{2},3)$, and
the
 orthorhombic superlattice peaks $(\frac{1}{2},\frac{1}{2},4)$,
$(\frac{1}{2},\frac{1}{2},2)$, and $(\frac{3}{2},\frac{3}{2},4)$, of
\lccco\ (a, b) and \lscco\ (c, d) as a function of temperature. \#1
and \#2 denote data from different experiments (see text). In (a) and
(b) errors are within point size. Inset: Full width at half maximum
of magnetic Bragg and orthorhombic superlattice peaks in \lscco\ as a
function of temperature. Spectrometer resolution indicated by dotted
line.} \label{fig4}
\end{figure}

Fig.~\ref{fig4} shows the intensity of representative magnetic and
structural superlattice peaks, after subtraction of the background,
as a function of temperature. First we will focus on \lccco , which
we have studied twice. In the first experiment (\#1), the intensity
of the magnetic ($\frac{1}{2},\frac{1}{2},3$) peak in
Fig.~\ref{fig4}(a) indicates N\'eel ordering above room temperature.
However, the relatively slow increase of the intensity below $T_N$
suggests that the phase transition is quite inhomogeneous. In
addition, two further transitions are observed: one at $\sim$175~K,
where the intensity starts to increase faster, and a second at 20~K,
where the intensity decreases. In a second experiment (\#2) performed
three
months later on the same crystal as well as on a second crystal,
$T_N$ had decreased to $\sim$175~K, though also in this case a very
weak peak could be traced up to 300~K. Moreover, the drop in
intensity below 20~K appeared to be weaker by a factor of 2-3. We
note that we have checked this for the same crystal, peak and
spectrometer configuration as in experiment \#1. However, a full
temperature scan was performed only on the ($\frac{1}{2},
\frac{3}{2}, 3$) peak after mounting this crystal in the ($h,k,2k$)
zone [Fig.~\ref{fig4}(a)]. Surprisingly, in this case the drop below
20~K is not observed at all. We explain this as follows: with the
crystal mounted in the ($h,k,2k$) zone the 2D scattering rods are
almost perpendicular to the horizontal scattering plane. Because of
the relaxed collimation in the vertical direction, scans through the
magnetic Bragg peak will also integrate some intensity from 2D
scattering. As the diffuse intensity increases at low temperature
(Sec.~\ref{diffuse}) it possibly compensates the decrease of the
magnetic Bragg peak below 20~K.

The clear decrease of the magnetic peak intensity below 20~K in
experiment \#1 strongly resembles findings for the spin-glass phase
in lightly Sr-doped \lsco\ with $x < 0.02$.~\cite{Matsuda02a} There,
the decrease of the commensurate magnetic Bragg peak is accompanied
by the appearance of intensity at peak positions consistent with
incommensurate stripe antiferromagnetism.~\cite{Matsuda02a}
Unfortunately, the apparent change of our sample prevented us from
taking similar measurements on \lccco\ in the same state as that
found in experiment \#1. In the second experiment no intensity at
incommensurate peak positions was observed (see Sec.~\ref{stripes}).
Although further experiments are needed to sort out why the sample
has changed, a very likely scenario is a redistribution of the oxygen
over time.

In contrast to the magnetic peaks, the temperature dependence of the
orthorhombic superlattice peaks of \lccco\ are nearly identical in
both experiments [Fig.~\ref{fig4}(b)]. Below the structural
transition at 175~K intensity increases and saturates at low
temperatures. It may, however, be important to notice that in
experiment \#1 the intensity slightly decreases below 20~K. We assume
that this decrease is caused by the magnetic contribution to the
$(\frac{1}{2},\frac{1}{2},4)$ peak, which is about 15-20\%.

Quite remarkably, the temperature profiles of magnetic and
superstructure peaks in experiment \#2 are almost identical [cf.
Fig.~\ref{fig4}(a) and (b)], indicating a magneto-elastic coupling.
The same similarity is observed for the \lscco\ crystal in
Fig.~\ref{fig4}(c) and (d), with the major differences being that the
transition occurs at around 125~K and the order is short-range. The
lower transition temperature and much weaker peak intensity most
likely follows from the higher hole and oxygen content than in \lccco
. We mention that similar data as for our \lscco\ crystal were
obtained by Ulrich~et al. on a \ulrich\ crystal with short-range
order.~\cite{Ulrich02a} However, our results on \lccco\ show that the
coupling of orthorhombic strain and antiferromagnetic order also
exists in samples with low charge carrier concentration and
long-range order (see discussion).
\begin{figure}[t]
\center{\includegraphics[width=1\columnwidth,angle=0,clip]{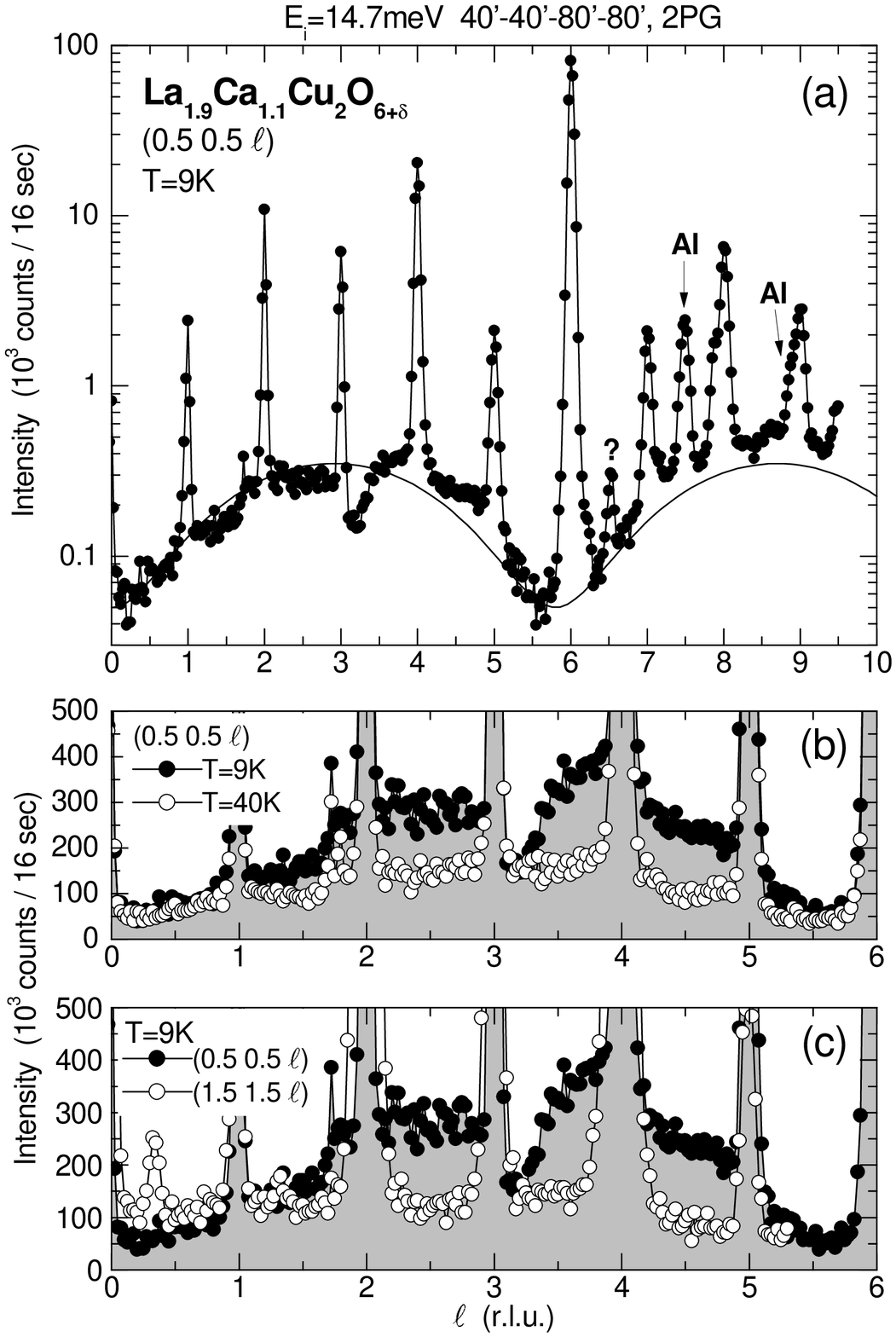}}
\caption[]{Elastic $l$-scans for \lccco . (a) along ${\bf
Q}=(\frac{1}{2},\frac{1}{2},l)$ at 9~K (logarithmic intensity scale).
The peak at $l=7.46$ as well as the left shoulder of the peak at
$l=9$ originate from Al powder ring reflections. (b) along ${\bf
Q}=(\frac{1}{2},\frac{1}{2},l)$ at 9~K and 40~K. (c) along ${\bf
Q}=(\frac{1}{2},\frac{1}{2},l)$ and $(\frac{3}{2},\frac{3}{2},l)$ at
9~K. The origin of the peaks at $l=6.5$ in (a) and at $l=0.35$ in (c)
is not clear, but there seems to be no systematic appearance of
further unidentified peaks.} \label{fig5}
\end{figure}

The inset in Fig.~\ref{fig4}(d) shows the full width at half maximum
of the magnetic and the orthorhombic in-plane peak width in
reciprocal lattice units of $a^*=2\pi/a$. The contribution of the
spectrometer resolution to the line width at the lowest temperature
amounts to 5\%. While the orthorhombic superlattice peak could be
analyzed up to 160~K, the magnetic peak became too weak for $T>70$~K.
Nevertheless, it is obvious that up to 70~K the magnetic peak width
is nearly the same as for the structural peak, and shows a similar
temperature dependence. For a comparison of the width of the
orthorhombic superlattice peaks in both of the crystals, we refer to
Sec.~\ref{xrays}, where we show high-resolution X-ray diffraction
results.
\begin{figure}[t]
\center{\includegraphics[width=1\columnwidth,angle=270,clip]{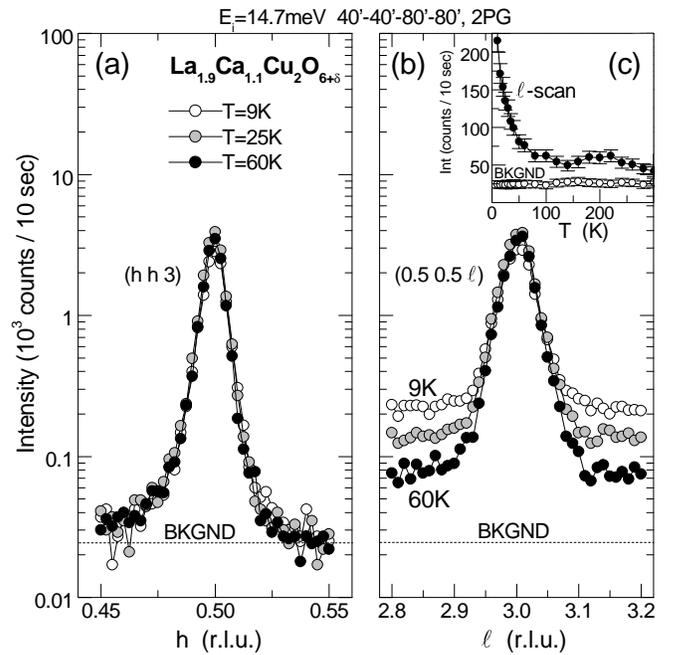}}
\caption[]{Elastic scans through the magnetic Bragg peak \ohoht\ of
\lccco\ at different temperatures. (a) $h$-scans. (b) $l$-scans. (c)
background signal as well as diffuse scattering intensity extracted
from $h$- and $l$-scans.} \label{fig6}
\end{figure}
\subsubsection{Diffuse scattering}
\label{diffuse}
A closer look at the data of \lccco\ reveals significant intensity
from elastic 2D scattering between the Bragg peaks (Fig.~\ref{fig5}).
At low $|{\bf Q}|$ it originates largely from magnetic scattering, as
will be discussed in the next paragraph. The long $l$-scan in
Fig.~\ref{fig5}(a), which is the same as in Fig.~\ref{fig3}(a), shows
that the 2D magnetic scattering intensity is modulated sinusoidally.
[Note that the dip in the data at $l=3.2$ was not reproducible, e.g.,
in an identical scan in Fig.~\ref{fig6}(b) it is absent.] The
modulation results from the antiferromagnetic coupling between the
$\rm CuO_2$ planes within a bilayer and is proportional to the square
of the magnetic structure factor $g({\bf q})$ for the acoustic spin
wave mode:~\cite{Tranquada89a, Shamoto93a, Dai01a}
\begin{equation}
g={\rm sin}(\pi z l)
\end{equation}
where $z$ is the ratio between the intra-bilayer distance $d$ of the
$\rm CuO_2$ planes and the lattice parameter $c$ (see
Fig.~\ref{fig1}). For lightly Ca- or Sr-doped \lcco\ the
literature~\cite{Shaked93a,Ulrich02a} gives a value of $z \simeq
0.17$, resulting in the solid line in Fig.~\ref{fig5}, which is in
good agreement with the modulation of the data. The first maximum of
the magnetic structure factor is reached for $l=2.9$, which is why
magnetic Bragg peaks are studied best at $l=3$. Note that for a
detailed description of the scattering intensity one has to include
the magnetic form factor and the spin structure, as well as the
spectrometer resolution, which we have neglected in this qualitative
discussion.~\cite{Tranquada89a} The magnetic form factor is a slowly
varying function and causes a decrease of intensity with increasing
$|{\bf Q}|$.~\cite{Shamoto93a} On the other hand, assuming a
collinear spin structure with the spins lying within the $\rm CuO_2$
planes (as in other cuprates), we expect an increase of intensity
with increasing $l$ associated with an increase of the magnetic
interaction vector $\bf S_\perp = \hat{Q} \times (S \times \hat{Q})$,
where $\bf S$ is the Cu spin and $\bf
\hat{Q}=Q/|Q|$.~\cite{Shirane02a} The $l$ dependence of $\bf S_\perp$
most likely explains why in Fig.~\ref{fig5}(a) the 2D scattering
intensity around the second maximum at $l=8.7$ is higher than around
the first maximum at $l=2.9$.~\cite{spinstructure}

\begin{figure}[t]
\center{\includegraphics[width=0.55\columnwidth,angle=0,clip]{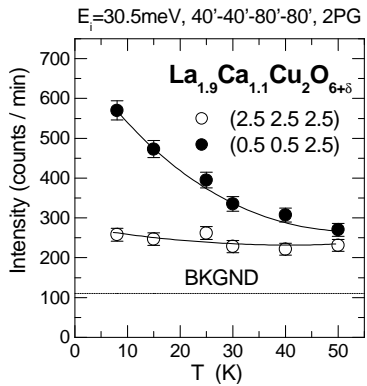}}
\caption[]{Elastic diffuse scattering intensity at
$(\frac{1}{2},\frac{1}{2},\frac{5}{2})$ and
$(\frac{5}{2},\frac{5}{2},\frac{5}{2})$ in \lccco\ as a function of
temperature.} \label{fig7}
\end{figure}

The elastic 2D magnetic scattering intensity strongly decreases with
increasing temperature as can be seen from Fig.~\ref{fig5}(b) where
we show a comparison of $l$-scans at 9~K and 40~K. A more detailed
temperature dependence was extracted from scans through the magnetic
\ohoht\ peak in Fig.~\ref{fig6}(a) and (b). In $h$-scans, the
intensity in the tails at any temperature quickly approaches the
background, while in $l$-scans it stays above the background due to
2D scattering. When increasing the temperature up to $\sim$100~K, the
2D scattering intensity drastically decreases, but stays above
background up to room temperature [Fig.~\ref{fig6}(c)].

It is important to notice that the temperature dependence of the
diffuse scattering intensity in Fig.~\ref{fig6}(c) is different from
that of the magnetic Bragg peaks in Fig.~\ref{fig4}(a). Instead it is
more similar to the temperature dependence of the short-range
magnetic order in \lscco\ in Fig.~\ref{fig4}(c). The increase of the
diffuse intensity is particularly steep at temperatures where the
decrease of the magnetic Bragg peak was observed, indicating that
intensity is transferred from the Bragg peaks to the 2D scattering
rods [cf. Fig.~\ref{fig6}(c) and \ref{fig4}(a)]. A similar transfer
of intensity at low temperatures was observed in \lsco\ with $x<0.02$
(Ref.~\onlinecite{Matsuda02a}) and in \ybcot\ (see Fig.~13 and 14 in
Ref.~\onlinecite{Tranquada89a}). Therefore, we assume that it is
associated with the freezing of spin fluctuations, which leads to the
formation of the spin glass phase at low temperatures.

To examine the origin of the elastic 2D scattering, we have studied
its $|{\bf Q}|$ dependence in \lccco . As is shown in
Fig.~\ref{fig5}(c), the diffuse scattering intensity in a scan along
${\bf Q} = (\frac{3}{2},\frac{3}{2},l)$ is significantly lower than
in a scan along ${\bf Q}=(\frac{1}{2},\frac{1}{2},l)$, indicating
that at low $|{\bf Q}|$ it results largely from 2D magnetic
scattering. On the other hand, $h$-scans along ${\bf Q} =
(h,h,\frac{5}{2})$ through the 2D scattering rods at $h=\frac{1}{2}$
and $h=\frac{5}{2}$ reveal that some of the intensity results from 2D
nuclear scattering. In Fig.~\ref{fig7}, where we show intensities,
one can see that the intensity at ${\bf Q} =
(\frac{1}{2},\frac{1}{2},\frac{5}{2})$ decreases with increasing
temperature, while at ${\bf Q} =
(\frac{5}{2},\frac{5}{2},\frac{5}{2})$ it is temperature independent,
indicating the dominance of 2D nuclear scattering for higher $|{\bf
Q}|$.

\begin{figure}[t]
\center{\includegraphics[width=0.54\columnwidth,angle=270,clip]{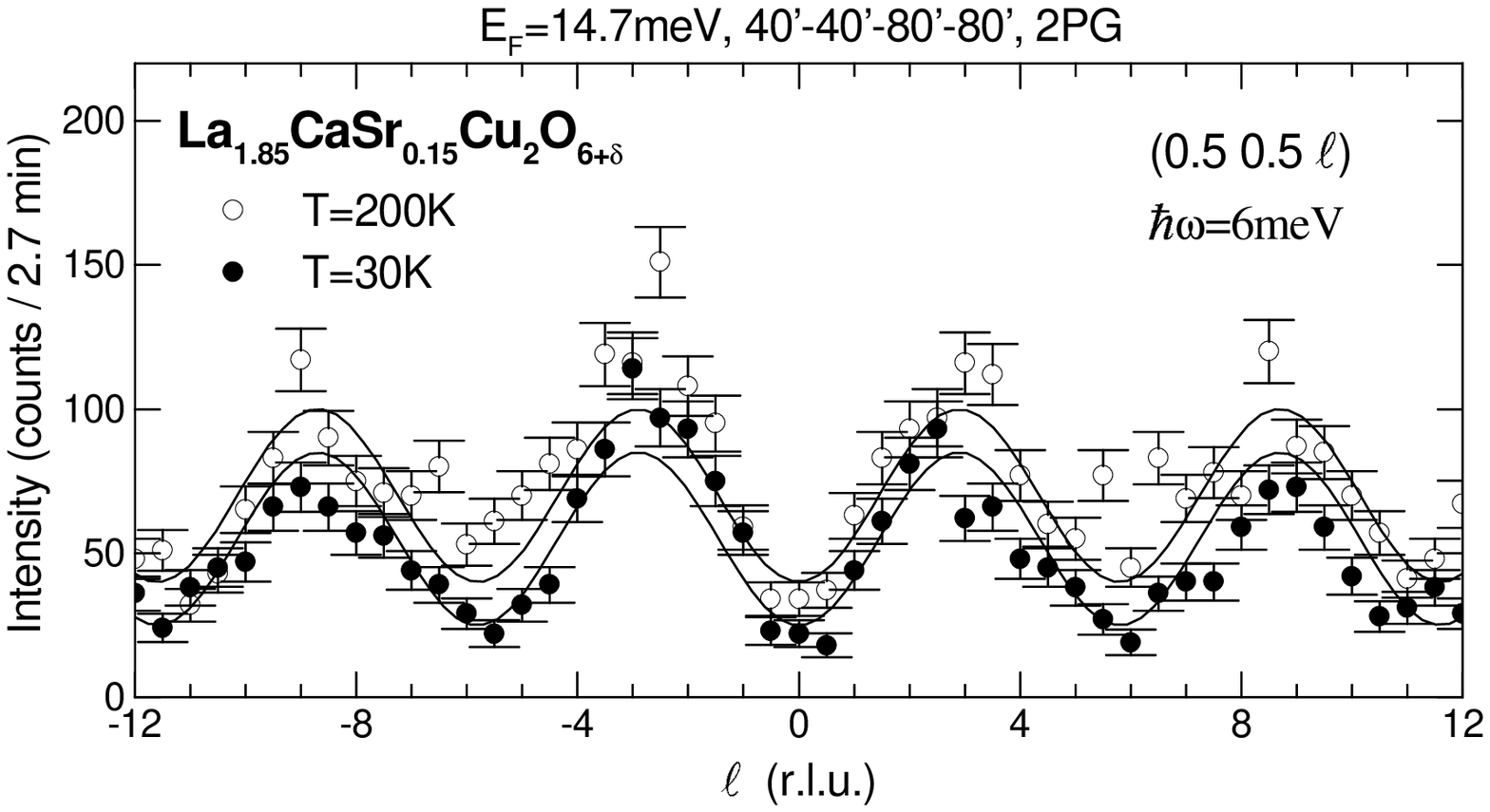}}
\caption[]{Inelastic scans with $\hbar\omega =6$~meV along ${\bf Q}
=(\frac{1}{2},\frac{1}{2},l)$ for \lscco .} \label{fig8}
\end{figure}

In \lscco\ the elastic 2D scattering intensity at 1.7~K is
significantly weaker than in \lccco\ at 9~K. It is modulated
sinusoidally along $l$ and decreases rapidly with increasing
temperature. The period is the same as for the Ca-doped crystal in
Fig.~\ref{fig5}. In \lscco\ the modulation is particularly pronounced
in inelastic scans, as is shown in Fig.~\ref{fig8} for scans along
${\bf Q} =(\frac{1}{2},\frac{1}{2},l)$ at 30~K and 200~K and an
energy transfer of $\hbar\omega =6$~meV. Note that this underlines
the magnetic nature of the diffuse scattering, as at finite energies
no significant contribution from diffuse nuclear scattering is
expected. From Fig.~\ref{fig8} it is obvious that the modulation of
the inelastic scattering intensity far above and below the magnetic
and structural transition at $\sim$ 125~K does not differ
significantly, indicating that in the paramagnetic phase the
intra-bilayer antiferromagnetic correlations are still strong. We
mention, that a similar structure factor modulation was observed in
\ybco .~\cite{Tranquada89a, Shamoto93a} However, there the relative
bilayer spacing amounts to $z=0.28$, so that the first maximum is
reached at $l=1.8$, in comparison to $l=2.9$ in \lcco . Furthermore,
in \ybco\ it was observed that at $l=-1.8$ the intensity is larger
than at $l=+1.8$, which is due to the so called focusing effect of
the spectrometer resolution.~\cite{Tranquada89a} At $l=+1.8$ the long
axis of the resolution ellipsoid has a significant angle with the 2D
scattering rod, therefore integrating less intensity than at
$l=-1.8$, where the long axis and the rod are approximately parallel.
For \lscco\ in Fig.~\ref{fig8} the intensity at $l= -2.9$ is only
slightly larger than at $l= +2.9$, which is due to the fact that the
2D scattering rod is relatively broad in $h$ and $k$, which
diminishes the focussing effect. Intensity clearly decreases with
increasing $l$ due to the magnetic form factor.

\subsubsection{Search for spin stripes} \label{stripes}
In the presence of antiferromagnetic stripe correlations similar to
those in \lsco\ we would expect intensity at incommensurate peak
positions displaced by ${\bf q}_s$ from the magnetic Bragg peaks at
${\bf Q}_{AF}$. Different sets of incommensurate peaks are possible,
depending on whether the direction of stripes is parallel to the
in-plane Cu-O bonds [${\bf q}_s = (\pm \epsilon,0,0)$ or $(0,\pm
\epsilon,0)$] or diagonal [${\bf q}_s = (\pm \sqrt{2}\epsilon,\pm
\sqrt{2}\epsilon,0)$ or $(\pm \sqrt{2}\epsilon,\mp
\sqrt{2}\epsilon,0)$].~\cite{Tranquada95a,Yamada98a,Fujita02a} Since
due to the magnetic structure factor the magnetic intensity is
maximum near $l = 3$, the crystals were mounted in the $(h,k,2k)$
zone, to scan these incommensurate positions around ${\bf Q}_{AF}=
(\frac{1}{2},\frac{3}{2},3)$ [see Fig.~\ref{fig2}(b)]. For both
stoichiometries, elastic scans at $T=8$~K indicate commensurate
antiferromagnetism. For the Sr-doped crystal the collimation was set
to $40'$-$40'$-$80'$-$80'$, and for the Ca-doped crystal to
$10'$-$20'$-$20'$-$20'$. Inelastic scans, performed on \lscco\ at
$T=8$~K, show commensurate antiferromagnetism, also. These scans were
performed at an energy transfer of $\hbar\omega = 3$~meV and a final
energy of 14.7~meV. The \lccco\ crystal was then mounted in the
$(h,3k,7k)$ zone to perform elastic scans at $T=8$~K through the 2D
scattering rod at ${\bf Q}=(\frac{1}{2},\frac{3}{2}, \frac{7}{2})$,
with the collimation set to $40'$-$40'$-$80'$-$80'$ (this type of
scans is not indicated in Fig.~\ref{fig2}).~\cite{ALpeak} The scans
show that the 2D magnetic scattering is commensurate, as well.

\begin{figure}[t]
\center{\includegraphics[width=0.95\columnwidth,angle=0,clip]{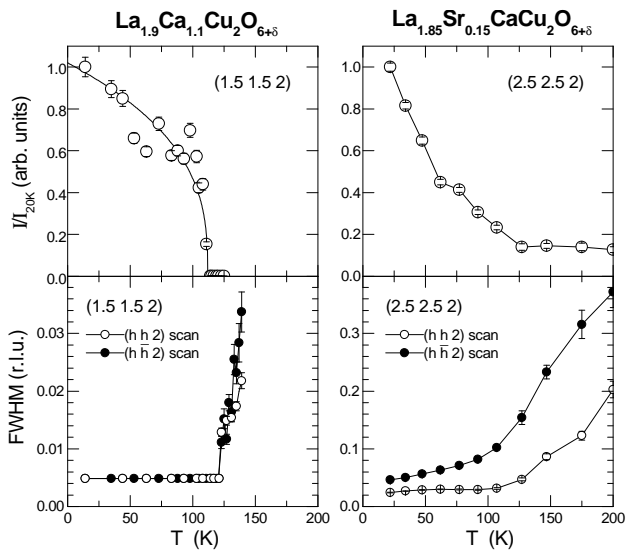}}
\caption[]{Results from X-ray diffraction: Normalized integrated
intensity and full width at half maximum of the orthorhombic
superlattice peak $(\frac{5}{2},\frac{5}{2},2)$ in \lccco\ (left) and
\lscco\ (right) as a function of temperature. Solid line in top left
graph is a guide to the eye.} \label{fig9}
\end{figure}

\subsection{X-ray Diffraction} \label{xrays}
In Fig.~\ref{fig9} we show single crystal X-ray diffraction data on
the orthorhombic superlattice peak. The polished surface of each
crystal was mounted so that $h$-scans along ${\bf Q}=(h,h,l)$ as well
as $(h,\overline{h},l)$ could be performed with high resolution.
Interestingly, in \lccco\ the structural phase transition is much
sharper and takes place at a lower temperature ($\sim$115~K) than in
the neutron scattering experiment ($\sim$170~K) [Fig.~\ref{fig4}(a)].
Below the transition the line width is limited by the spectrometer
resolution but significantly increases above the transition. For
\lscco\ we observe a broad continuous transition with a critical
temperature around $125$~K, very similar to the neutron diffraction
data in Fig.~\ref{fig4}. The peak width is about one order of
magnitude larger than that in \lccco , and its temperature dependence
is in good agreement with our results from neutron diffraction.
Although above the transition temperature intensity is very small,
with X-rays the superlattice peak could be followed up to 200~K
revealing a strong increase of the peak width. In both crystals the
superlattice peaks are sharper for $h$-scans along ${\bf Q}=(h,h,l)$
than in the perpendicular direction ${\bf Q}=(h,\overline{h},l)$.
According to Ulrich et al. the structural distortion in the
orthorhombic phase consists of a displacement of the apical oxygen
O(2) along [110] (cf. Fig.~\ref{fig1}). Hence, our data seem to
indicate a larger domain size along the displacement direction than
in the perpendicular in-plane direction.

\subsection{Static Susceptibility}
\label{magnetism}
\begin{figure}[t]
\center{\includegraphics[width=0.9\columnwidth,angle=0,clip]{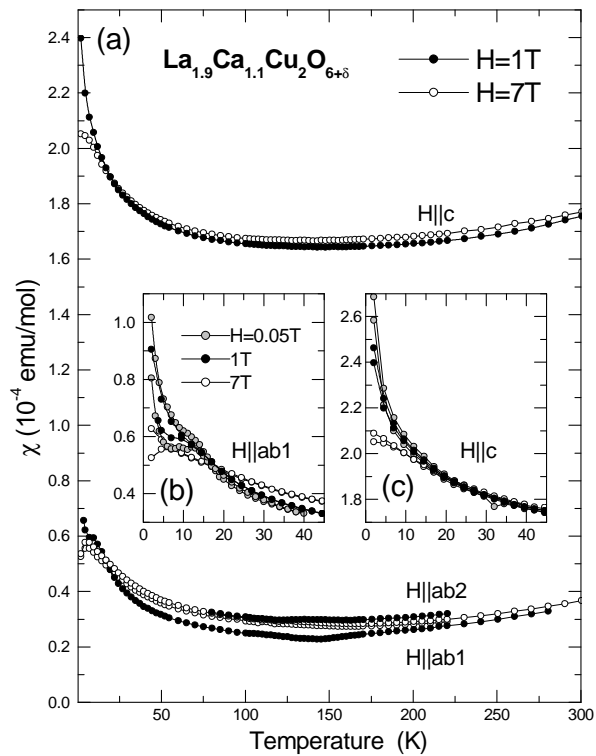}}
\caption[]{Static magnetic \sus\ of \lccco\ as a function of
temperature and different directions of the applied field. (a) Zero
field cooled data. (b), (c) Zero field cooled and field cooled data
at low temperatures.} \label{fig10}
\end{figure}

To compare our results from diffraction with macroscopic magnetic
properties we have measured the static \sus\ $\chi$ of small pieces
of the crystals. First we will focus on the \sus\ of \lccco .
Figure~\ref{fig10}(a) shows its $\chi$ in $H=1$~T and $7$~T for $H$
parallel to the $c$-axis as well as parallel to the two in-plane
directions [110] and [1\={1}0], which are parallel to the
orthorhombic $a$ and $b$ axis. Because of twinning these two in-plane
directions are mixed, which is why we call them $ab1$ and $ab2$. From
room temperature the \sus\ slightly decreases with decreasing $T$ for
all three directions, which is typical for a cuprate S=1/2,
2D-Heisenberg antiferromagnet with a superexchange coupling constant
$J$ of the order of 0.1~eV.~\cite{Okabe88,Johnston89} The Curie-type
increase at low temperatures usually follows from magnetic impurities
such as defect Cu spins. The difference between $H
\parallel c$ and $H \parallel ab$ of $\sim 1.4 \times
10^{-4}$~emu/mol is largely due to the Cu Van Vleck magnetism, i.e.,
$\sim 0.7 \times 10^{-4}$~emu/mol per $\rm CuO_2$ plane, which is in
good agreement with
literature.~\cite{Allgeier93,Terasaki92,Lavrov01a} For $H
\parallel c$, no anomaly is observed at the antiferromagnetic ordering temperature, and
data for 1~T and 7~T are nearly identical. However, for magnetic field
parallel to the $\rm CuO_2$ planes we find a weak field dependence as well
as a small anisotropy between $H\parallel ab1$ and $ab2$. The anisotropy
indicates that the crystal is partially detwinned. At a field of 7~T,
however, the anisotropy has vanished.

\begin{figure}[t]
\center{\includegraphics[width=1\columnwidth,angle=0,clip]{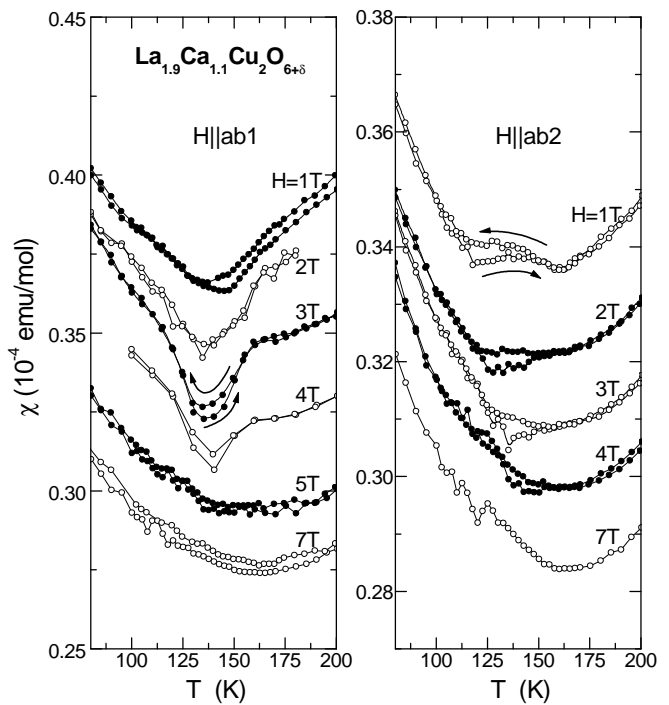}}
\caption[]{Static magnetic \sus\ of \lccco\ for different magnetic
fields as a function of increasing and decreasing temperature. (a)
for $H\parallel ab1$ and (b) for $H\parallel ab2$. Curves shifted for
clarity.} \label{fig11}
\end{figure}

Figures~\ref{fig10}(b) and (c) show zero-field-cooled and
field-cooled data for different $H$ at low temperatures. While for
$H\parallel c$ data are nearly reversible, clear differences, as well
as a hump, are observed for $H\parallel ab1$ and $H\parallel ab2$
(not shown) indicating the occurrence of a spin-glass transition at
$\sim$13~K. This is in contrast to \lsco\ where the spin-glass
transition is visible for $H\parallel c$,
also.~\cite{Chou95,Wakimoto00b}

\begin{figure}[t]
\center{\includegraphics[width=0.82\columnwidth,angle=0,clip]{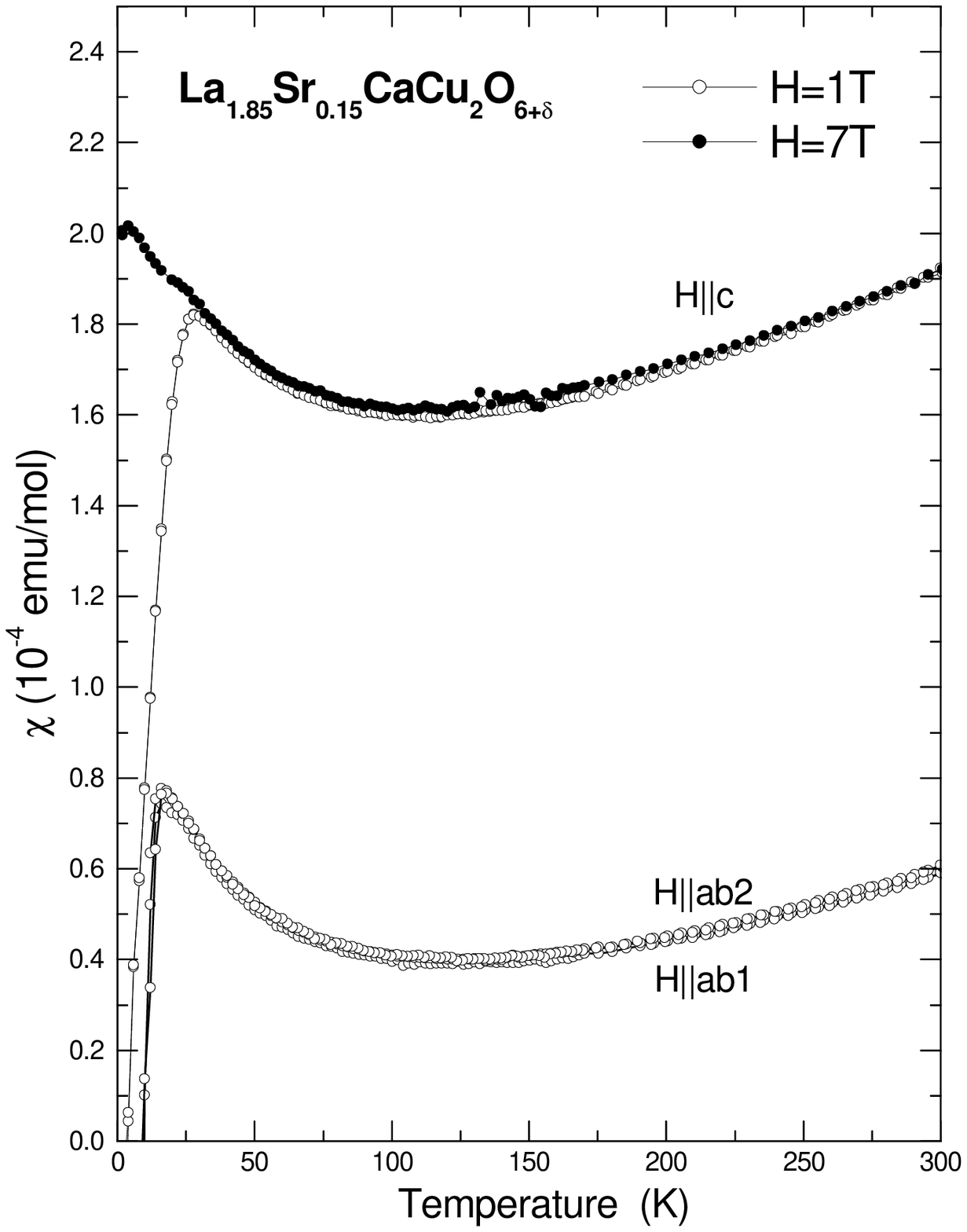}}
\caption[]{Static magnetic \sus\ of \lscco\ as a function of
temperature and different directions of the applied field.}
\label{fig12}
\end{figure}

The fact that in \lsco\ both the N\'eel and the spin-glass transition
are clearly visible for $H\parallel c$ is connected to a weak
out-of-plane \DM\ spin canting, which follows from the rotation of
the $\rm CuO_6$ octahedra in the orthorhombic
phase.~\cite{Thio88,Huecker02b} As mentioned earlier, in \lccco\ the
orthorhombic distortion consists primarily of a displacement of the
apical oxygen O(2) along [110], while the basal oxygens O(1) seem to
stay in the $\rm CuO_2$ plane.~\cite{Ulrich02a} Hence, the absence of
a signature at the N\'eel and the spin-glass transition for
$H\parallel c$ in \lccco\ might be interpreted as evidence for the
absence of an out-of-plane \DM\ spin canting. However, it can also
mean that the \DM\ spin canting cancels due to a strong
antiferromagnetic coupling between the $\rm CuO_2$ planes of a
bilayer. Since it is reasonable to assume that, similar to \ybco ,
this interlayer coupling within a bilayer is orders of magnitude
stronger than the inter-bilayer coupling, it will most likely be very
difficult to test for the presence of a \DM\ spin canting by means of
static magnetization measurements.~\cite{Thio88,Huecker02b}

As is shown in Fig.~\ref{fig11} for magnetic field $H=1$~T parallel
to the $\rm CuO_2$ planes, further anomalies are observed at
$\sim$115~K and $\sim$160~K. In particular, for $H\parallel ab1$
these two temperatures border a dip-like anomaly, and for the
perpendicular direction $H
\parallel ab2$ a hump. In both cases the anomalies are suppressed by a
magnetic field of the order of $4-5$~T and at 7~T the \sus\ is
isotropic (cf. Fig.~\ref{fig10}). We mention that for $H=1$~T applied
parallel to [100] no anomalies are observed; however, they can be
induced at fields of $3-5$~T and then disappear again at 7~T.
Therefore, we assume that below $T_N$ spins are either parallel to
[110] or [1\={1}0] and the observed transitions as a function of
field are due to a spin flop. The relatively low field scale
indicates an in-plane spin wave gap of the order of 0.5~meV. This gap
energy is much smaller than in \lco\ where the in-plane gap was
associated with the \DM\ spin canting, which might confirm that the
spin canting in \lcco\ is indeed small.~\cite{Keimer93,Thio94} For
comparison, in \lco\ the spin flop field amounts to 10-15~T, and in
tetragonal \scoc\ with perfectly coplanar spin structure it is only
0.7~T.~\cite{Thio90,Vaknin90}

Interestingly, the two critical temperatures of 115~K and 160~K are
nearly identical with the structural transition temperatures, as
observed by XRD and ND, respectively (cf. Fig.~\ref{fig4} and
\ref{fig9}). Considering our diffraction data alone, possible
interpretations are (i) that we have indeed observed two distinct
transitions, (ii) that surface effects are important (XRD), (iii)
that a inhomogeneous distribution of the oxygen content exists. The
fact that both anomalies show up in the macroscopic \sus\ indicates
that they involve large volume fractions, making surface effects less
likely. In the case of an inhomogeneous oxygen distribution we would
expect one broad transition rather than two relatively sharp ones. On
the other hand, for ND, probing the entire crystal, the transition is
in fact relatively broad, while for XRD, probing a relatively small
area of the sample, the transition is sharp. Hence, diffraction data
are not inconsistent with an inhomogeneous oxygen distribution. In
this case the anomaly at 115~K might correspond to the $T_N$ of the
hole rich surface region and the anomaly at 160~K the $T_N$ of the
bulk, which has a slightly lower average hole concentration. On the
other hand, the \sus\ data in Fig.~\ref{fig11}(a) are difficult to
understand in terms of a sequence of two transitions of the same kind
differing only in their critical temperatures, since at $\sim$160~K
$\chi$ first decreases and then again increases at $\sim$115~K.
Further experiments are needed to clarify the origin of this
intriguing sequence of transitions.

Now let us turn to $\chi(T)$ of \lscco , which is shown in
Fig.~\ref{fig12}. In a field of $H=1$~T the onset of
superconductivity is observed at $\sim$25~K for $H\parallel c$ and at
$\sim$15~K for $H\parallel ab$. At a field of 7~T applied $\parallel
c$ superconductivity is suppressed completely. Above $T_c$ the data
look similar to that of \lccco\ in Fig.~\ref{fig10}. However, no
anomaly is observed at the transition to short-range
antiferromagnetic order at $\sim 125$~K. Interestingly, the increase
of $\chi$ at high temperatures is steeper than for the Ca-doped
sample. This is a clear indication for a shift of the high
temperature maximum of the Heisenberg spin susceptibility
$\chi_{2DHAF}$ to lower temperatures, which is consistent with the
fact that the hole content in \lscco\ is larger and therefore the
spin stiffness is smaller than in \lccco .~\cite{Johnston89}

\subsection{Discussion}
\label{discussion}

The two major results in this study are (i) the similarity between
the thermal evolution of the orthorhombic distortion and
antiferromagnetic order for short as well as long-range ordered
samples, and (ii) the fact that the antiferromagnetic correlations
are commensurate. For a proper interpretation one has to consider the
hole concentration, which is known to strongly change the electronic
ground state of high-temperature superconductors.~\cite{Orenstein00a}
As mentioned in the introduction, we have estimated the degree of
hole doping in our crystals from their optical conductivity in
Ref.~\onlinecite{Wang03a}. Both crystals show considerable optical
weight at low wave numbers, indicating a significant concentration of
holes. In fact, when compared to optical data on \lsco , the hole
content appears to be close to the nominal value $p=x/2$, which
suggests that the oxygen content $\delta$ in our crystals is close to
zero.~\cite{Uchida91} Good agreement is also found with optical data
in Ref.~\onlinecite{Shibata93a}. In particular, the room temperature
optical data of our \lccco\ crystal grown at 1~atm $\rm O_2$ are
nearly identical to that of a grown-in-air \shibata\
crystal.~\cite{Wang03a,Shibata93a} Corresponding data for our \lscco\
crystal grown at 11~atm indicate a higher hole concentration,
comparable to \shibata\ annealed at 20-100~atm.~\cite{Shibata93a}
According to Ref.~\onlinecite{Shibata93a}, \shibata\ is not
superconducting when grown in air nor after annealing at 20 atm $\rm
O_2$, while annealing at 100 atm results in a $T_c$ of 13~K in the
resistivity. The presence of weak superconductivity in our Sr-doped
crystal and its absence in the Ca-doped one is compatible with this,
and shows that the Sr-doped crystal is just on the borderline to the
superconducting phase.

If we assume that the hole contents in the crystals are close to the
nominal values, then the relatively high N\'eel temperature in
\lccco\ and onset temperature for short-range order in \lscco\ are
surprising, when compared to \lsco . However, in the bilayer system
\ybccos\ long-range order is suppressed at a higher hole content and
spin-glass transition temperatures are higher than in \lsco , as
well.~\cite{Niedermayer98} This indicates that in bilayer systems
antiferromagnetic order in general is more stable, which is due to
the relatively strong coupling between the $\rm CuO_2$ planes of a
bilayer.~\cite{Tranquada89a} In the case of $\rm La_{2-{\it
x}}(Ca,Sr)_{\it x}CaCu_2O_{6+{\it \delta}}$ it seems, however, that
additional degrees of freedom may play a vital role, as will be
discussed in the next paragraphs.

%

The intriguing similarity between the temperature dependence of the
magnetic Bragg peak and the orthorhombic superlattice peak fuels the
idea of a significant magneto-elastic coupling. In particular, we
expect that the magnetic inter-bilayer coupling depends on the
orthorhombic strain. In the tetragonal high-temperature phase the
magnetic inter-bilayer coupling is strongly frustrated, because of
the centro symmetric arrangement of spins in adjacent bilayers (cf.
Fig.~\ref{fig1}). The orthorhombic distortion at low temperature
lifts the frustration, thereby triggering 3D antiferromagnetic order.
A similar situation exists in the single layer compounds \lco\ and
\scoc . While \lco\ becomes orthorhombic below 525~K and orders
antiferromagnetically at $T_N=325$~K, \scoc\ with $T_N=250$~K stays
tetragonal.~\cite{Boeni88,Johnston89,Vaknin90,Greven95} It is
believed that the orthorhombic strain in \lco\ is one reason for the
higher N\'eel temperature, since it lifts the frustration of the
interlayer coupling.~\cite{Vaknin90,Xue88,Greven95,Suh95} We mention
that early $\mu$SR results on \lcco\ and \lsrco\ indicate N\'eel
temperatures higher than 250~K for the undoped
system.~\cite{Ansaldo92a} On the other hand, in
Ref.~\onlinecite{Ulrich02a} the structural transition temperature in
a \lcco\ powder sample was determined to be $\sim$175~K. This means
that 3D magnetic order can exist in the tetragonal high temperature
phase.~\cite{undoped} However, in our hole doped \lccco\ crystal,
where $T_N$ is considerably reduced from its maximum value, the
structural transition seems to be essential for the stabilization of
3D antiferromagnetic order.

In the case of the \lscco\ crystal the magneto-elastic coupling seems
to still be active. This conclusion is  based on a similar
temperature dependence for the peak intensity and correlation length
of the structural distortion and antiferromagnetic correlations (cf.
Fig.~\ref{fig4}). Since in Ref.~\onlinecite{Ulrich02a} nearly
identical results were found for a \ulrich\ crystal, we assume that
this behavior is intrinsic for samples grown at low oxygen pressure,
which are relatively close to the superconducting phase, but not yet
good superconductors. However, one has to mention that while neutron
diffraction experiments on \ulrich\ indicate short-range magnetic
order below $\sim 100$~K, muon spin rotation measurements find static
order only below 10~K.~\cite{Ulrich02a} This shows that for $10~{\rm
K} < T < 100~{\rm K}$ the magnetic correlations are not yet truly
static. In this respect both \lsrcco\ crystals (this work and
Ref.~\onlinecite{Ulrich02a}) qualitatively show a similar behavior as
\lsco\ in the spin-glass phase.~\cite{Keimer92b,Matsuda00a,Hiraka01a}
However, there are major differences which we believe signal a
significant magneto-elastic coupling. First, the ``elastic" magnetic
intensities we observe in \lscco\ below the structural transition at
$\sim 125$~K are much larger than in \lsco\ at a comparable hole
doping close to the metal insulator transition.~\cite{Hiraka01a} The
onset temperature is much higher and seems to correlate with the
structural transition. Moreover, in \lscco\ magnetic correlations are
commensurate, while they are incommensurate in \lsco
.~\cite{Hiraka01a,Wakimoto00a,Ulrich02a,Fujita02a}

Since in \lccco\ the inter-bilayer exchange seems to couple very
clearly to the orthorhombic lattice distortion, it is reasonable to
assume that this coupling is at least one reason why in \lscco\ the
freezing of the short-range magnetic order, as observed by neutron
diffraction, reflects the temperature dependence of the structural
distortion. In contrast, in \lsco\ it is believed that the spin
freezing occurs upon the localization of the holes at low
temperatures.~\cite{Takagi92b,Chou93,Gooding98a} It is very likely
that hole localization is relevant in \lscco , also. But whether the
structural transition into the orthorhombic phase expedites the hole
localization, is unclear. In Ref.~\onlinecite{Ulrich02a} it is argued
that both structural as well as magnetic disorder follows from the
localization of holes. However, the results on \lccco\ seem to show
quite clearly that it is the orthorhombic structure which stabilizes
the antiferromagnetic order. Therefore, we assume that the coupling
of the lattice to charge degrees of freedom is not responsible for
the structural disorder observed in \lscco . Spin freezing, on the
contrary, not only is coupled to charge localization -- as is the
case in \lsco\ -- but also to the short-range lattice distortions.
The electronic disorder, possibly induced or enhanced by structural
disorder, might be responsible for the absence of incommensurate
magnetic correlations at low temperatures. As is well known,
structural disorder in this bilayer system is caused by the mixed
occupation of the metal sites M(1) and M(2), the doping with Sr and
Ca as well as by interstitial oxygen O(3) and defects in the oxygen
matrix (cf. Fig.~\ref{fig1}).~\cite{Kinoshita92a,Ohyama95a,Shaked93a}

\section{Conclusion}
\label{conclusion}
Structural and magnetic properties of the bilayer cuprates \lccco\
and \lscco\ were studied by means of neutron and X-ray scattering as
well as static \sus\ measurements. We observe an intimate connection
between antiferromagnetic order and the structural lattice
distortion, which we explain in terms of an orthorhombic
strain-induced magnetic inter-bilayer coupling. While in the Ca-doped
crystal structural distortions and the Cu spins are long-range
ordered, the Sr-doped crystal exhibits short-range order as well as
weak superconductivity. In both crystals antiferromagnetic
correlations are commensurate, i.e., no direct evidence for
incommensurate spin stripes is found. However, in \lccco\ the
temperature dependence of the magnetic diffuse scattering as well as
the static susceptibility indicate the presence of a spin-glass
phase. In \lsco\ this type of phase was shown to exhibit a
short-range spin stripe
order.~\cite{Wakimoto99a,Fujita02a,Matsuda02a} The reasons for the
absence of incommensurate spin correlations in particular in \lscco\
are not understood, but might result from an inhomogeneity of the
charge and spin density, possibly induced by the short-range ordered
structural distortions in the orthorhombic phase.

\begin{acknowledgments}

We acknowledge support by the Office of Science, US Department of Energy
under Contract No. DE-AC02-98CH10886.

\end{acknowledgments}


\begin{thebibliography}{68}
\expandafter\ifx\csname
natexlab\endcsname\relax\def\natexlab#1{#1}\fi
\expandafter\ifx\csname bibnamefont\endcsname\relax
  \def\bibnamefont#1{#1}\fi
\expandafter\ifx\csname bibfnamefont\endcsname\relax
  \def\bibfnamefont#1{#1}\fi
\expandafter\ifx\csname citenamefont\endcsname\relax
  \def\citenamefont#1{#1}\fi
\expandafter\ifx\csname url\endcsname\relax
  \def\url#1{\texttt{#1}}\fi
\expandafter\ifx\csname urlprefix\endcsname\relax\def\urlprefix{URL
}\fi \providecommand{\bibinfo}[2]{#2}
\providecommand{\eprint}[2][]{\url{#2}}

\bibitem[{\citenamefont{Cava}(1990)}]{Cava90b}
\bibinfo{author}{\bibfnamefont{R.~J.} \bibnamefont{Cava}},
  \bibinfo{journal}{Science} \textbf{\bibinfo{volume}{247}},
  \bibinfo{pages}{656} (\bibinfo{year}{1990}).

\bibitem[{\citenamefont{Cava et~al.}(1990)\citenamefont{Cava, Batlogg, van\
  Dover, Krajewski, Waszczak, Fleming, Jr., Jr., Marsh, James
  et~al.}}]{Cava90c}
\bibinfo{author}{\bibfnamefont{R.~J.} \bibnamefont{Cava}},
  \bibinfo{author}{\bibfnamefont{B.}~\bibnamefont{Batlogg}},
  \bibinfo{author}{\bibfnamefont{R.~B.} \bibnamefont{van\ Dover}},
  \bibinfo{author}{\bibfnamefont{J.~J.} \bibnamefont{Krajewski}},
  \bibinfo{author}{\bibfnamefont{J.~V.} \bibnamefont{Waszczak}},
  \bibinfo{author}{\bibfnamefont{R.~M.} \bibnamefont{Fleming}},
  \bibinfo{author}{\bibfnamefont{W.~F.~P.} \bibnamefont{Jr.}},
  \bibinfo{author}{\bibfnamefont{L.~W.~R.} \bibnamefont{Jr.}},
  \bibinfo{author}{\bibfnamefont{P.}~\bibnamefont{Marsh}},
  \bibinfo{author}{\bibfnamefont{A.~C. W.~P.} \bibnamefont{James}},
  \bibnamefont{et~al.}, \bibinfo{journal}{Nature}
  \textbf{\bibinfo{volume}{345}}, \bibinfo{pages}{602} (\bibinfo{year}{1990}).

\bibitem[{\citenamefont{Kinoshita and Yamada}(1992)}]{Kinoshita92b}
\bibinfo{author}{\bibfnamefont{K.}~\bibnamefont{Kinoshita}} \bibnamefont{and}
  \bibinfo{author}{\bibfnamefont{T.}~\bibnamefont{Yamada}},
  \bibinfo{journal}{Phys.\ Rev. B} \textbf{\bibinfo{volume}{46}},
  \bibinfo{pages}{9116} (\bibinfo{year}{1992}).

\bibitem[{\citenamefont{Klamut et~al.}(2000)\citenamefont{Klamut, Dabrowski,
  Dybzinski, and Bukowski}}]{Klamut00a}
\bibinfo{author}{\bibfnamefont{P.~W.} \bibnamefont{Klamut}},
  \bibinfo{author}{\bibfnamefont{B.}~\bibnamefont{Dabrowski}},
  \bibinfo{author}{\bibfnamefont{R.}~\bibnamefont{Dybzinski}},
  \bibnamefont{and} \bibinfo{author}{\bibfnamefont{Z.}~\bibnamefont{Bukowski}},
  \bibinfo{journal}{J.\ Appl.\ Phys.} \textbf{\bibinfo{volume}{87}},
  \bibinfo{pages}{5558} (\bibinfo{year}{2000}).

\bibitem[{\citenamefont{Pavarini et~al.}(2001)\citenamefont{Pavarini, Dasgupta,
  Saha-Dasgupta, Jepsen, and Andersen}}]{Pavarini01a}
\bibinfo{author}{\bibfnamefont{E.}~\bibnamefont{Pavarini}},
  \bibinfo{author}{\bibfnamefont{I.}~\bibnamefont{Dasgupta}},
  \bibinfo{author}{\bibfnamefont{T.}~\bibnamefont{Saha-Dasgupta}},
  \bibinfo{author}{\bibfnamefont{O.}~\bibnamefont{Jepsen}}, \bibnamefont{and}
  \bibinfo{author}{\bibfnamefont{O.~K.} \bibnamefont{Andersen}},
  \bibinfo{journal}{Phys.\ Rev.\ Lett.} \textbf{\bibinfo{volume}{87}},
  \bibinfo{pages}{047003} (\bibinfo{year}{2001}).

\bibitem[{\citenamefont{Shaked et~al.}(1993)\citenamefont{Shaked, Jorgensen,
  Hunter, Hitterman, Kinoshita, Izumi, and Kamiyama}}]{Shaked93a}
\bibinfo{author}{\bibfnamefont{H.}~\bibnamefont{Shaked}},
  \bibinfo{author}{\bibfnamefont{J.~D.} \bibnamefont{Jorgensen}},
  \bibinfo{author}{\bibfnamefont{B.~A.} \bibnamefont{Hunter}},
  \bibinfo{author}{\bibfnamefont{R.~L.} \bibnamefont{Hitterman}},
  \bibinfo{author}{\bibfnamefont{K.}~\bibnamefont{Kinoshita}},
  \bibinfo{author}{\bibfnamefont{F.}~\bibnamefont{Izumi}}, \bibnamefont{and}
  \bibinfo{author}{\bibfnamefont{T.}~\bibnamefont{Kamiyama}},
  \bibinfo{journal}{Phys.\ Rev.\ B} \textbf{\bibinfo{volume}{48}},
  \bibinfo{pages}{12941} (\bibinfo{year}{1993}).

\bibitem[{\citenamefont{Orenstein and Millis}(2000)}]{Orenstein00a}
\bibinfo{author}{\bibfnamefont{J.}~\bibnamefont{Orenstein}} \bibnamefont{and}
  \bibinfo{author}{\bibfnamefont{A.~J.} \bibnamefont{Millis}},
  \bibinfo{journal}{Science} \textbf{\bibinfo{volume}{288}},
  \bibinfo{pages}{468} (\bibinfo{year}{2000}).

\bibitem[{\citenamefont{Cheong et~al.}(1991)\citenamefont{Cheong, Aeppli,
  Mason, Mook, Hayden, Canfield, Fisk, Clausen, and Martinez}}]{Cheong91}
\bibinfo{author}{\bibfnamefont{S.-W.} \bibnamefont{Cheong}},
  \bibinfo{author}{\bibfnamefont{G.}~\bibnamefont{Aeppli}},
  \bibinfo{author}{\bibfnamefont{T.~E.} \bibnamefont{Mason}},
  \bibinfo{author}{\bibfnamefont{H.}~\bibnamefont{Mook}},
  \bibinfo{author}{\bibfnamefont{S.~M.} \bibnamefont{Hayden}},
  \bibinfo{author}{\bibfnamefont{P.~C.} \bibnamefont{Canfield}},
  \bibinfo{author}{\bibfnamefont{Z.}~\bibnamefont{Fisk}},
  \bibinfo{author}{\bibfnamefont{K.~N.} \bibnamefont{Clausen}},
  \bibnamefont{and} \bibinfo{author}{\bibfnamefont{J.~L.}
  \bibnamefont{Martinez}}, \bibinfo{journal}{Phys.\ Rev.\ Lett.}
  \textbf{\bibinfo{volume}{67}}, \bibinfo{pages}{1791} (\bibinfo{year}{1991}).

\bibitem[{\citenamefont{Mason et~al.}(1992)\citenamefont{Mason, Aeppli, and
  Mook}}]{Mason92}
\bibinfo{author}{\bibfnamefont{T.~E.} \bibnamefont{Mason}},
  \bibinfo{author}{\bibfnamefont{G.}~\bibnamefont{Aeppli}}, \bibnamefont{and}
  \bibinfo{author}{\bibfnamefont{H.~A.} \bibnamefont{Mook}},
  \bibinfo{journal}{Phys.\ Rev.\ Lett.} \textbf{\bibinfo{volume}{68}},
  \bibinfo{pages}{1414} (\bibinfo{year}{1992}).

\bibitem[{\citenamefont{Hayden et~al.}(1992)\citenamefont{Hayden, Lander,
  Zarestky, Brown, Stassis, Metcalf, and Honig}}]{Hayden92aN}
\bibinfo{author}{\bibfnamefont{S.~M.} \bibnamefont{Hayden}},
  \bibinfo{author}{\bibfnamefont{G.~H.} \bibnamefont{Lander}},
  \bibinfo{author}{\bibfnamefont{J.}~\bibnamefont{Zarestky}},
  \bibinfo{author}{\bibfnamefont{P.~J.} \bibnamefont{Brown}},
  \bibinfo{author}{\bibfnamefont{C.}~\bibnamefont{Stassis}},
  \bibinfo{author}{\bibfnamefont{P.}~\bibnamefont{Metcalf}}, \bibnamefont{and}
  \bibinfo{author}{\bibfnamefont{J.~M.} \bibnamefont{Honig}},
  \bibinfo{journal}{Phys. Rev. Lett.} \textbf{\bibinfo{volume}{68}},
  \bibinfo{pages}{1061} (\bibinfo{year}{1992}).

\bibitem[{\citenamefont{Tranquada et~al.}(1995)\citenamefont{Tranquada,
  Sternlieb, Axe, Nakamura, and Uchida}}]{Tranquada95a}
\bibinfo{author}{\bibfnamefont{J.~M.} \bibnamefont{Tranquada}},
  \bibinfo{author}{\bibfnamefont{B.~J.} \bibnamefont{Sternlieb}},
  \bibinfo{author}{\bibfnamefont{J.~D.} \bibnamefont{Axe}},
  \bibinfo{author}{\bibfnamefont{Y.}~\bibnamefont{Nakamura}}, \bibnamefont{and}
  \bibinfo{author}{\bibfnamefont{S.}~\bibnamefont{Uchida}},
  \bibinfo{journal}{Nature} \textbf{\bibinfo{volume}{375}},
  \bibinfo{pages}{561} (\bibinfo{year}{1995}).

\bibitem[{\citenamefont{Mook et~al.}(2002)\citenamefont{Mook, Dai, and
  Do\u{g}an}}]{Mook02a}
\bibinfo{author}{\bibfnamefont{H.~A.} \bibnamefont{Mook}},
  \bibinfo{author}{\bibfnamefont{P.}~\bibnamefont{Dai}}, \bibnamefont{and}
  \bibinfo{author}{\bibfnamefont{F.}~\bibnamefont{Do\u{g}an}},
  \bibinfo{journal}{Phys.\ Rev.\ Lett.} \textbf{\bibinfo{volume}{88}},
  \bibinfo{pages}{97004} (\bibinfo{year}{2002}).

\bibitem[{\citenamefont{Dai et~al.}(2001)\citenamefont{Dai, Mook, Hunt, and
  Dogan}}]{Dai01a}
\bibinfo{author}{\bibfnamefont{P.}~\bibnamefont{Dai}},
  \bibinfo{author}{\bibfnamefont{H.~A.} \bibnamefont{Mook}},
  \bibinfo{author}{\bibfnamefont{R.~D.} \bibnamefont{Hunt}}, \bibnamefont{and}
  \bibinfo{author}{\bibfnamefont{F.}~\bibnamefont{Dogan}},
  \bibinfo{journal}{Phys.\ Rev.\ B} \textbf{\bibinfo{volume}{63}},
  \bibinfo{pages}{54525} (\bibinfo{year}{2001}).

\bibitem[{\citenamefont{Zaanen}(2000)}]{Zaanen00a}
\bibinfo{author}{\bibfnamefont{J.}~\bibnamefont{Zaanen}},
  \bibinfo{journal}{Nature} \textbf{\bibinfo{volume}{404}},
  \bibinfo{pages}{714} (\bibinfo{year}{2000}).

\bibitem[{\citenamefont{Kivelson et~al.}(2003)\citenamefont{Kivelson, Bindloss,
  Fradkin, Oganesyan, Tranquada, Kapitulnik, and Howald}}]{Kivelson03a}
\bibinfo{author}{\bibfnamefont{S.~A.} \bibnamefont{Kivelson}},
  \bibinfo{author}{\bibfnamefont{I.~P.} \bibnamefont{Bindloss}},
  \bibinfo{author}{\bibfnamefont{E.}~\bibnamefont{Fradkin}},
  \bibinfo{author}{\bibfnamefont{V.}~\bibnamefont{Oganesyan}},
  \bibinfo{author}{\bibfnamefont{J.~M.} \bibnamefont{Tranquada}},
  \bibinfo{author}{\bibfnamefont{A.}~\bibnamefont{Kapitulnik}},
  \bibnamefont{and} \bibinfo{author}{\bibfnamefont{C.}~\bibnamefont{Howald}},
  \bibinfo{journal}{Rev.\ Mod.\ Phys.} \textbf{\bibinfo{volume}{75}},
  \bibinfo{pages}{1201} (\bibinfo{year}{2003}).

\bibitem[{\citenamefont{Tranquada et~al.}(2004)\citenamefont{Tranquada, Woo,
  Perring, Goka, Gu, Xu, Fujita, and Yamada}}]{Tranquada04a}
\bibinfo{author}{\bibfnamefont{J.~M.} \bibnamefont{Tranquada}},
  \bibinfo{author}{\bibfnamefont{H.}~\bibnamefont{Woo}},
  \bibinfo{author}{\bibfnamefont{T.~G.} \bibnamefont{Perring}},
  \bibinfo{author}{\bibfnamefont{H.}~\bibnamefont{Goka}},
  \bibinfo{author}{\bibfnamefont{G.~D.} \bibnamefont{Gu}},
  \bibinfo{author}{\bibfnamefont{G.}~\bibnamefont{Xu}},
  \bibinfo{author}{\bibfnamefont{M.}~\bibnamefont{Fujita}}, \bibnamefont{and}
  \bibinfo{author}{\bibfnamefont{K.}~\bibnamefont{Yamada}},
  \bibinfo{journal}{Nature} \textbf{\bibinfo{volume}{429}},
  \bibinfo{pages}{534} (\bibinfo{year}{2004}).

\bibitem[{\citenamefont{Hayden et~al.}(2004)\citenamefont{Hayden, Mook, Dai,
  Perring, and Do\u{g}an}}]{Hayden04a}
\bibinfo{author}{\bibfnamefont{S.~M.} \bibnamefont{Hayden}},
  \bibinfo{author}{\bibfnamefont{H.~A.} \bibnamefont{Mook}},
  \bibinfo{author}{\bibfnamefont{P.}~\bibnamefont{Dai}},
  \bibinfo{author}{\bibfnamefont{T.~G.} \bibnamefont{Perring}},
  \bibnamefont{and}
  \bibinfo{author}{\bibfnamefont{F.}~\bibnamefont{Do\u{g}an}},
  \bibinfo{journal}{Nature} \textbf{\bibinfo{volume}{429}},
  \bibinfo{pages}{531} (\bibinfo{year}{2004}).

\bibitem[{\citenamefont{Ulrich et~al.}(2002)\citenamefont{Ulrich, Kondo,
  Reehuis, He, Bernhard, Niedermayer, Bour\'ee, Bourges, Ohl, R{\o}nnow
  et~al.}}]{Ulrich02a}
\bibinfo{author}{\bibfnamefont{C.}~\bibnamefont{Ulrich}},
  \bibinfo{author}{\bibfnamefont{S.}~\bibnamefont{Kondo}},
  \bibinfo{author}{\bibfnamefont{M.}~\bibnamefont{Reehuis}},
  \bibinfo{author}{\bibfnamefont{H.}~\bibnamefont{He}},
  \bibinfo{author}{\bibfnamefont{C.}~\bibnamefont{Bernhard}},
  \bibinfo{author}{\bibfnamefont{C.}~\bibnamefont{Niedermayer}},
  \bibinfo{author}{\bibfnamefont{F.}~\bibnamefont{Bour\'ee}},
  \bibinfo{author}{\bibfnamefont{P.}~\bibnamefont{Bourges}},
  \bibinfo{author}{\bibfnamefont{M.}~\bibnamefont{Ohl}},
  \bibinfo{author}{\bibfnamefont{H.}~\bibnamefont{R{\o}nnow}},
  \bibnamefont{et~al.}, \bibinfo{journal}{Phys.\ Rev.\ B}
  \textbf{\bibinfo{volume}{65}}, \bibinfo{pages}{220507}
  (\bibinfo{year}{2002}).

\bibitem[{\citenamefont{Wakimoto
  et~al.}(2000{\natexlab{a}})\citenamefont{Wakimoto, Birgeneau, Kastner, Erwin,
  Gehring, Lee, Fujita, Yamada, Endoh, Hirota et~al.}}]{Wakimoto00a}
\bibinfo{author}{\bibfnamefont{S.}~\bibnamefont{Wakimoto}},
  \bibinfo{author}{\bibfnamefont{R.~J.} \bibnamefont{Birgeneau}},
  \bibinfo{author}{\bibfnamefont{M.~A.} \bibnamefont{Kastner}},
  \bibinfo{author}{\bibfnamefont{Y.~S. L.~R.} \bibnamefont{Erwin}},
  \bibinfo{author}{\bibfnamefont{P.~M.} \bibnamefont{Gehring}},
  \bibinfo{author}{\bibfnamefont{S.~H.} \bibnamefont{Lee}},
  \bibinfo{author}{\bibfnamefont{M.}~\bibnamefont{Fujita}},
  \bibinfo{author}{\bibfnamefont{K.}~\bibnamefont{Yamada}},
  \bibinfo{author}{\bibfnamefont{Y.}~\bibnamefont{Endoh}},
  \bibinfo{author}{\bibfnamefont{K.}~\bibnamefont{Hirota}},
  \bibnamefont{et~al.}, \bibinfo{journal}{Phys.\ Rev.\ B}
  \textbf{\bibinfo{volume}{61}}, \bibinfo{pages}{3699}
  (\bibinfo{year}{2000}{\natexlab{a}}).

\bibitem[{\citenamefont{Yamada et~al.}(1998)\citenamefont{Yamada, Lee,
  Kurahashi, Wada, Wakimoto, Ueki, Kimura, Endoh, Hosoya, Shirane
  et~al.}}]{Yamada98a}
\bibinfo{author}{\bibfnamefont{K.}~\bibnamefont{Yamada}},
  \bibinfo{author}{\bibfnamefont{C.~H.} \bibnamefont{Lee}},
  \bibinfo{author}{\bibfnamefont{K.}~\bibnamefont{Kurahashi}},
  \bibinfo{author}{\bibfnamefont{J.}~\bibnamefont{Wada}},
  \bibinfo{author}{\bibfnamefont{S.}~\bibnamefont{Wakimoto}},
  \bibinfo{author}{\bibfnamefont{S.}~\bibnamefont{Ueki}},
  \bibinfo{author}{\bibfnamefont{H.}~\bibnamefont{Kimura}},
  \bibinfo{author}{\bibfnamefont{Y.}~\bibnamefont{Endoh}},
  \bibinfo{author}{\bibfnamefont{S.}~\bibnamefont{Hosoya}},
  \bibinfo{author}{\bibfnamefont{G.}~\bibnamefont{Shirane}},
  \bibnamefont{et~al.}, \bibinfo{journal}{Phys.\ Rev.\ B}
  \textbf{\bibinfo{volume}{57}}, \bibinfo{pages}{6165}
  (\bibinfo{year}{1998}).

\bibitem[{oxy()}]{oxygen}
  \bibinfo{note}{There is evidence that even in the case of $\delta= 0$
  some O(3) sites are occupied, which is compensated by a corresponding number
  of defects on the O(1) or O(2)
  sites~\cite{Izumi89a,Grasmeder90a,Lightfoot90a, Kinoshita92b,Shaked93a}}.

\bibitem[{\citenamefont{Ohyama et~al.}(1995)\citenamefont{Ohyama, Ohashi,
  Fukunaga, Ikawa, Izumi, and Tanaka}}]{Ohyama95a}
\bibinfo{author}{\bibfnamefont{T.}~\bibnamefont{Ohyama}},
  \bibinfo{author}{\bibfnamefont{N.}~\bibnamefont{Ohashi}},
  \bibinfo{author}{\bibfnamefont{O.}~\bibnamefont{Fukunaga}},
  \bibinfo{author}{\bibfnamefont{H.}~\bibnamefont{Ikawa}},
  \bibinfo{author}{\bibfnamefont{F.}~\bibnamefont{Izumi}}, \bibnamefont{and}
  \bibinfo{author}{\bibfnamefont{J.}~\bibnamefont{Tanaka}},
  \bibinfo{journal}{Physica} \textbf{\bibinfo{volume}{C~249}},
  \bibinfo{pages}{293} (\bibinfo{year}{1995}).

\bibitem[{\citenamefont{Deng et~al.}(1999)\citenamefont{Deng, Dong, Chen, Wu,
  Jia, Shen, and Zhao}}]{Deng99b}
\bibinfo{author}{\bibfnamefont{H.}~\bibnamefont{Deng}},
  \bibinfo{author}{\bibfnamefont{C.}~\bibnamefont{Dong}},
  \bibinfo{author}{\bibfnamefont{H.}~\bibnamefont{Chen}},
  \bibinfo{author}{\bibfnamefont{F.}~\bibnamefont{Wu}},
  \bibinfo{author}{\bibfnamefont{S.~L.} \bibnamefont{Jia}},
  \bibinfo{author}{\bibfnamefont{J.~C.} \bibnamefont{Shen}}, \bibnamefont{and}
  \bibinfo{author}{\bibfnamefont{Z.~X.} \bibnamefont{Zhao}},
  \bibinfo{journal}{Physica} \textbf{\bibinfo{volume}{C~313}},
  \bibinfo{pages}{285} (\bibinfo{year}{1999}).

\bibitem[{\citenamefont{Izumi et~al.}(1989)\citenamefont{Izumi,
  Takayama-Muromachi, Nakai, and Asano}}]{Izumi89a}
\bibinfo{author}{\bibfnamefont{F.}~\bibnamefont{Izumi}},
  \bibinfo{author}{\bibfnamefont{E.}~\bibnamefont{Takayama-Muromachi}},
  \bibinfo{author}{\bibfnamefont{Y.}~\bibnamefont{Nakai}}, \bibnamefont{and}
  \bibinfo{author}{\bibfnamefont{H.}~\bibnamefont{Asano}},
  \bibinfo{journal}{Physica} \textbf{\bibinfo{volume}{C~157}},
  \bibinfo{pages}{89} (\bibinfo{year}{1989}).

\bibitem[{\citenamefont{Grasmeder and Weller}(1990)}]{Grasmeder90a}
\bibinfo{author}{\bibfnamefont{J.~R.} \bibnamefont{Grasmeder}}
  \bibnamefont{and} \bibinfo{author}{\bibfnamefont{M.~T.}
  \bibnamefont{Weller}}, \bibinfo{journal}{J.\ Solid State \ Chem.}
  \textbf{\bibinfo{volume}{85}}, \bibinfo{pages}{88} (\bibinfo{year}{1990}).

\bibitem[{\citenamefont{Lightfoot et~al.}(1990)\citenamefont{Lightfoot, Pei,
  Jorgensen, Tang, Manthiram, and Goodenough}}]{Lightfoot90a}
\bibinfo{author}{\bibfnamefont{P.}~\bibnamefont{Lightfoot}},
  \bibinfo{author}{\bibfnamefont{S.}~\bibnamefont{Pei}},
  \bibinfo{author}{\bibfnamefont{J.~D.} \bibnamefont{Jorgensen}},
  \bibinfo{author}{\bibfnamefont{X.-X.} \bibnamefont{Tang}},
  \bibinfo{author}{\bibfnamefont{A.}~\bibnamefont{Manthiram}},
  \bibnamefont{and} \bibinfo{author}{\bibfnamefont{J.~B.}
  \bibnamefont{Goodenough}}, \bibinfo{journal}{Physica}
  \textbf{\bibinfo{volume}{C~169}}, \bibinfo{pages}{464}
  (\bibinfo{year}{1990}).

\bibitem[{\citenamefont{Wang et~al.}(2003)\citenamefont{Wang, Zheng, Feng, Gu,
  Homes, Tranquada, Gaulin, and Timusk}}]{Wang03a}
\bibinfo{author}{\bibfnamefont{N.~L.} \bibnamefont{Wang}},
  \bibinfo{author}{\bibfnamefont{P.}~\bibnamefont{Zheng}},
  \bibinfo{author}{\bibfnamefont{T.}~\bibnamefont{Feng}},
  \bibinfo{author}{\bibfnamefont{G.~D.} \bibnamefont{Gu}},
  \bibinfo{author}{\bibfnamefont{C.~C.} \bibnamefont{Homes}},
  \bibinfo{author}{\bibfnamefont{J.~M.} \bibnamefont{Tranquada}},
  \bibinfo{author}{\bibfnamefont{B.~D.} \bibnamefont{Gaulin}},
  \bibnamefont{and} \bibinfo{author}{\bibfnamefont{T.}~\bibnamefont{Timusk}},
  \bibinfo{journal}{Phys.\ Rev.\ B} \textbf{\bibinfo{volume}{67}},
  \bibinfo{pages}{134526} (\bibinfo{year}{2003}).

\bibitem[{\citenamefont{Freltoft et~al.}(1987)\citenamefont{Freltoft, Fischer,
  Shirane, Moncton, Sinha, Vaknin, Remeika, Cooper, and Harshman}}]{Freltoft87}
\bibinfo{author}{\bibfnamefont{T.}~\bibnamefont{Freltoft}},
  \bibinfo{author}{\bibfnamefont{J.~E.} \bibnamefont{Fischer}},
  \bibinfo{author}{\bibfnamefont{G.}~\bibnamefont{Shirane}},
  \bibinfo{author}{\bibfnamefont{D.~E.} \bibnamefont{Moncton}},
  \bibinfo{author}{\bibfnamefont{S.~K.} \bibnamefont{Sinha}},
  \bibinfo{author}{\bibfnamefont{D.}~\bibnamefont{Vaknin}},
  \bibinfo{author}{\bibfnamefont{J.~P.} \bibnamefont{Remeika}},
  \bibinfo{author}{\bibfnamefont{A.~S.} \bibnamefont{Cooper}},
  \bibnamefont{and} \bibinfo{author}{\bibfnamefont{D.}~\bibnamefont{Harshman}},
  \bibinfo{journal}{Phys.\ Rev.\ B} \textbf{\bibinfo{volume}{36}},
  \bibinfo{pages}{R826} (\bibinfo{year}{1987}).

\bibitem[{\citenamefont{Tranquada et~al.}(1988)\citenamefont{Tranquada, Cox,
  Kunnmann, Moudden, Shirane, Suenaga, Zolliker, Vaknin, Sinha, Alvarez
  et~al.}}]{Tranquada88b}
\bibinfo{author}{\bibfnamefont{J.~M.} \bibnamefont{Tranquada}},
  \bibinfo{author}{\bibfnamefont{D.~E.} \bibnamefont{Cox}},
  \bibinfo{author}{\bibfnamefont{W.}~\bibnamefont{Kunnmann}},
  \bibinfo{author}{\bibfnamefont{H.}~\bibnamefont{Moudden}},
  \bibinfo{author}{\bibfnamefont{G.}~\bibnamefont{Shirane}},
  \bibinfo{author}{\bibfnamefont{M.}~\bibnamefont{Suenaga}},
  \bibinfo{author}{\bibfnamefont{P.}~\bibnamefont{Zolliker}},
  \bibinfo{author}{\bibfnamefont{D.}~\bibnamefont{Vaknin}},
  \bibinfo{author}{\bibfnamefont{S.~K.} \bibnamefont{Sinha}},
  \bibinfo{author}{\bibfnamefont{M.~S.} \bibnamefont{Alvarez}},
  \bibnamefont{et~al.}, \bibinfo{journal}{Phys.\ Rev.\ Lett.}
  \textbf{\bibinfo{volume}{60}}, \bibinfo{pages}{156} (\bibinfo{year}{1988}).

\bibitem[{\citenamefont{B\"oni et~al.}(1988)\citenamefont{B\"oni, Axe, Shirane,
  Birgeneau, Gabbe, Jenssen, Kastner, Peters, Picone, and Thurston}}]{Boeni88}
\bibinfo{author}{\bibfnamefont{P.}~\bibnamefont{B\"oni}},
  \bibinfo{author}{\bibfnamefont{J.~D.} \bibnamefont{Axe}},
  \bibinfo{author}{\bibfnamefont{G.}~\bibnamefont{Shirane}},
  \bibinfo{author}{\bibfnamefont{R.~J.} \bibnamefont{Birgeneau}},
  \bibinfo{author}{\bibfnamefont{D.~R.} \bibnamefont{Gabbe}},
  \bibinfo{author}{\bibfnamefont{H.~P.} \bibnamefont{Jenssen}},
  \bibinfo{author}{\bibfnamefont{M.~A.} \bibnamefont{Kastner}},
  \bibinfo{author}{\bibfnamefont{C.~J.} \bibnamefont{Peters}},
  \bibinfo{author}{\bibfnamefont{P.~J.} \bibnamefont{Picone}},
  \bibnamefont{and} \bibinfo{author}{\bibfnamefont{T.~R.}
  \bibnamefont{Thurston}}, \bibinfo{journal}{Phys.\ Rev.\ B}
  \textbf{\bibinfo{volume}{38}}, \bibinfo{pages}{185} (\bibinfo{year}{1988}).

\bibitem[{\citenamefont{Radaelli et~al.}(1994)\citenamefont{Radaelli, Hinks,
  Mitchell, Hunter, Wagner, Dabrowski, Vandervoort, Viswanathan, and
  Jorgensen}}]{Radaelli94}
\bibinfo{author}{\bibfnamefont{P.~G.} \bibnamefont{Radaelli}},
  \bibinfo{author}{\bibfnamefont{D.~G.} \bibnamefont{Hinks}},
  \bibinfo{author}{\bibfnamefont{A.~W.} \bibnamefont{Mitchell}},
  \bibinfo{author}{\bibfnamefont{B.~A.} \bibnamefont{Hunter}},
  \bibinfo{author}{\bibfnamefont{J.~L.} \bibnamefont{Wagner}},
  \bibinfo{author}{\bibfnamefont{B.}~\bibnamefont{Dabrowski}},
  \bibinfo{author}{\bibfnamefont{K.~G.} \bibnamefont{Vandervoort}},
  \bibinfo{author}{\bibfnamefont{H.~K.} \bibnamefont{Viswanathan}},
  \bibnamefont{and} \bibinfo{author}{\bibfnamefont{J.~D.}
  \bibnamefont{Jorgensen}}, \bibinfo{journal}{Phys.\ Rev.\ B}
  \textbf{\bibinfo{volume}{49}}, \bibinfo{pages}{4163}
  (\bibinfo{year}{1994}).

\bibitem[{cor()}]{correl}
  \bibinfo{note}{Correlation lengths were calculated from the peak width
  using $\xi_a=({\rm HWHM} \times a^*)^{-1}$ and $\xi_c=({\rm HWHM} \times
  c^*)^{-1}$ where HWHM is the half width at half maximum in reciprocal lattice
  units of $a^*=1.64~{\rm \AA} ^{-1}$ and $c^*=0.323~{\rm \AA} ^{-1}$,
  respectively.}

\bibitem[{\citenamefont{Matsuda et~al.}(2002)\citenamefont{Matsuda, Fujita,
  Yamada, Birgeneau, Endoh, and Shirane}}]{Matsuda02a}
\bibinfo{author}{\bibfnamefont{M.}~\bibnamefont{Matsuda}},
  \bibinfo{author}{\bibfnamefont{M.}~\bibnamefont{Fujita}},
  \bibinfo{author}{\bibfnamefont{K.}~\bibnamefont{Yamada}},
  \bibinfo{author}{\bibfnamefont{R.~J.} \bibnamefont{Birgeneau}},
  \bibinfo{author}{\bibfnamefont{Y.}~\bibnamefont{Endoh}}, \bibnamefont{and}
  \bibinfo{author}{\bibfnamefont{G.}~\bibnamefont{Shirane}},
  \bibinfo{journal}{Phys.\ Rev.\ B} \textbf{\bibinfo{volume}{65}},
  \bibinfo{pages}{134515} (\bibinfo{year}{2002}).

\bibitem[{\citenamefont{Tranquada et~al.}(1989)\citenamefont{Tranquada,
  Shirane, Keimer, Shamoto, and Sato}}]{Tranquada89a}
\bibinfo{author}{\bibfnamefont{J.~M.} \bibnamefont{Tranquada}},
  \bibinfo{author}{\bibfnamefont{G.}~\bibnamefont{Shirane}},
  \bibinfo{author}{\bibfnamefont{B.}~\bibnamefont{Keimer}},
  \bibinfo{author}{\bibfnamefont{S.}~\bibnamefont{Shamoto}}, \bibnamefont{and}
  \bibinfo{author}{\bibfnamefont{M.}~\bibnamefont{Sato}},
  \bibinfo{journal}{Phys.\ Rev.\ B} \textbf{\bibinfo{volume}{40}},
  \bibinfo{pages}{4503} (\bibinfo{year}{1989}).

\bibitem[{\citenamefont{Shamoto et~al.}(1993)\citenamefont{Shamoto, Sato,
  Tranquada, Sternlieb, and Shirane}}]{Shamoto93a}
\bibinfo{author}{\bibfnamefont{S.}~\bibnamefont{Shamoto}},
  \bibinfo{author}{\bibfnamefont{M.}~\bibnamefont{Sato}},
  \bibinfo{author}{\bibfnamefont{J.~M.} \bibnamefont{Tranquada}},
  \bibinfo{author}{\bibfnamefont{B.~J.} \bibnamefont{Sternlieb}},
  \bibnamefont{and} \bibinfo{author}{\bibfnamefont{G.}~\bibnamefont{Shirane}},
  \bibinfo{journal}{Phys.\ Rev.\ B} \textbf{\bibinfo{volume}{48}},
  \bibinfo{pages}{13817} (\bibinfo{year}{1993}).

\bibitem[{\citenamefont{Shirane et~al.}(2002)\citenamefont{Shirane, Shapiro,
  and Tranquada}}]{Shirane02a}
\bibinfo{author}{\bibfnamefont{G.}~\bibnamefont{Shirane}},
  \bibinfo{author}{\bibfnamefont{S.~M.} \bibnamefont{Shapiro}},
  \bibnamefont{and} \bibinfo{author}{\bibfnamefont{J.~M.}
  \bibnamefont{Tranquada}}, \emph{\bibinfo{title}{Neutron Scattering with a
  Triple-Axis Spectrometer}} (\bibinfo{publisher}{Cambridge University Press},
  \bibinfo{year}{2002}).

\bibitem[{spi()}]{spinstructure}
  \bibinfo{note}{In the case that the spins are lying in the $\rm CuO_2$
  planes, for ${\bf Q} \parallel {\bf c}$ all spins are perpendicular to ${\bf
  Q}$, while for ${\bf Q} \perp {\bf c}$ this is generally not the case,
  depending on the in-plane spin direction and the arrangement of the magnetic
  domains relative to $\bf Q$. As only the spin component perpendicular to
  ${\bf Q}$ contributes to the scattering amplitude, intensity usually
  increases with increasing $l$.}

\bibitem[{\citenamefont{Fujita et~al.}(2002)\citenamefont{Fujita, Yamada,
  Hiraka, Gehring, Lee, Wakimoto, and Shirane}}]{Fujita02a}
\bibinfo{author}{\bibfnamefont{M.}~\bibnamefont{Fujita}},
  \bibinfo{author}{\bibfnamefont{K.}~\bibnamefont{Yamada}},
  \bibinfo{author}{\bibfnamefont{H.}~\bibnamefont{Hiraka}},
  \bibinfo{author}{\bibfnamefont{P.~M.} \bibnamefont{Gehring}},
  \bibinfo{author}{\bibfnamefont{S.~H.} \bibnamefont{Lee}},
  \bibinfo{author}{\bibfnamefont{S.}~\bibnamefont{Wakimoto}}, \bibnamefont{and}
  \bibinfo{author}{\bibfnamefont{G.}~\bibnamefont{Shirane}},
  \bibinfo{journal}{Phys.\ Rev.\ B} \textbf{\bibinfo{volume}{65}},
  \bibinfo{pages}{64505} (\bibinfo{year}{2002}).

\bibitem[{ALp()}]{ALpeak}
 \bibinfo{note}{Similar scans through the 2D scattering rod at ${\bf
  Q}=(\frac{1}{2},\frac{3}{2},\frac{5}{2})$ [with the crystal mounted in the
  $(h,3k,5k)$ zone] are contaminated by Al-powder peaks.}

\bibitem[{\citenamefont{Okabe and Kikuchi}(1988)}]{Okabe88}
\bibinfo{author}{\bibfnamefont{Y.}~\bibnamefont{Okabe}} \bibnamefont{and}
  \bibinfo{author}{\bibfnamefont{M.}~\bibnamefont{Kikuchi}},
  \bibinfo{journal}{J.\ Phys.\ Soc.\ Japan} \textbf{\bibinfo{volume}{57}},
  \bibinfo{pages}{4351} (\bibinfo{year}{1988}).

\bibitem[{\citenamefont{Johnston}(1989)}]{Johnston89}
\bibinfo{author}{\bibfnamefont{D.~C.} \bibnamefont{Johnston}},
  \bibinfo{journal}{Phys.\ Rev.\ Lett.} \textbf{\bibinfo{volume}{62}},
  \bibinfo{pages}{957} (\bibinfo{year}{1989}).

\bibitem[{\citenamefont{Allgeier and Schilling}(1993)}]{Allgeier93}
\bibinfo{author}{\bibfnamefont{C.}~\bibnamefont{Allgeier}} \bibnamefont{and}
  \bibinfo{author}{\bibfnamefont{J.~S.} \bibnamefont{Schilling}},
  \bibinfo{journal}{Phys.\ Rev.\ B} \textbf{\bibinfo{volume}{48}},
  \bibinfo{pages}{9747} (\bibinfo{year}{1993}).

\bibitem[{\citenamefont{Terasaki et~al.}(1992)\citenamefont{Terasaki, Hase,
  Maeda, Uchinokura, Kimura, Kishio, Tanaka, and Kojima}}]{Terasaki92}
\bibinfo{author}{\bibfnamefont{I.}~\bibnamefont{Terasaki}},
  \bibinfo{author}{\bibfnamefont{M.}~\bibnamefont{Hase}},
  \bibinfo{author}{\bibfnamefont{A.}~\bibnamefont{Maeda}},
  \bibinfo{author}{\bibfnamefont{K.}~\bibnamefont{Uchinokura}},
  \bibinfo{author}{\bibfnamefont{T.}~\bibnamefont{Kimura}},
  \bibinfo{author}{\bibfnamefont{K.}~\bibnamefont{Kishio}},
  \bibinfo{author}{\bibfnamefont{I.}~\bibnamefont{Tanaka}}, \bibnamefont{and}
  \bibinfo{author}{\bibfnamefont{H.}~\bibnamefont{Kojima}},
  \bibinfo{journal}{Physica} \textbf{\bibinfo{volume}{C~193}},
  \bibinfo{pages}{365} (\bibinfo{year}{1992}).

\bibitem[{\citenamefont{Lavrov et~al.}(2001)\citenamefont{Lavrov, Ando, Komiya,
  and Tsukada}}]{Lavrov01a}
\bibinfo{author}{\bibfnamefont{A.~N.} \bibnamefont{Lavrov}},
  \bibinfo{author}{\bibfnamefont{Y.}~\bibnamefont{Ando}},
  \bibinfo{author}{\bibfnamefont{S.}~\bibnamefont{Komiya}}, \bibnamefont{and}
  \bibinfo{author}{\bibfnamefont{I.}~\bibnamefont{Tsukada}},
  \bibinfo{journal}{Phys.\ Rev.\ Lett.} \textbf{\bibinfo{volume}{87}},
  \bibinfo{pages}{017007} (\bibinfo{year}{2001}).

\bibitem[{\citenamefont{Chou et~al.}(1995)\citenamefont{Chou, Belk, Kastner,
  Birgeneau, and Aharony}}]{Chou95}
\bibinfo{author}{\bibfnamefont{F.~C.} \bibnamefont{Chou}},
  \bibinfo{author}{\bibfnamefont{N.~R.} \bibnamefont{Belk}},
  \bibinfo{author}{\bibfnamefont{M.~A.} \bibnamefont{Kastner}},
  \bibinfo{author}{\bibfnamefont{R.~J.} \bibnamefont{Birgeneau}},
  \bibnamefont{and} \bibinfo{author}{\bibfnamefont{A.}~\bibnamefont{Aharony}},
  \bibinfo{journal}{Phys.\ Rev.\ Lett.} \textbf{\bibinfo{volume}{75}},
  \bibinfo{pages}{2204} (\bibinfo{year}{1995}).

\bibitem[{\citenamefont{Wakimoto
  et~al.}(2000{\natexlab{b}})\citenamefont{Wakimoto, Ueki, and
  Endoh}}]{Wakimoto00b}
\bibinfo{author}{\bibfnamefont{S.}~\bibnamefont{Wakimoto}},
  \bibinfo{author}{\bibfnamefont{S.}~\bibnamefont{Ueki}}, \bibnamefont{and}
  \bibinfo{author}{\bibfnamefont{Y.}~\bibnamefont{Endoh}},
  \bibinfo{journal}{Phys.\ Rev.\ B} \textbf{\bibinfo{volume}{62}},
  \bibinfo{pages}{3547} (\bibinfo{year}{2000}{\natexlab{b}}).

\bibitem[{\citenamefont{Thio et~al.}(1988)\citenamefont{Thio, Thurston, Preyer,
  Picone, Kastner, Jenssen, Gabbe, Chen, Birgeneau, and Aharony}}]{Thio88}
\bibinfo{author}{\bibfnamefont{T.}~\bibnamefont{Thio}},
  \bibinfo{author}{\bibfnamefont{T.~R.} \bibnamefont{Thurston}},
  \bibinfo{author}{\bibfnamefont{N.~W.} \bibnamefont{Preyer}},
  \bibinfo{author}{\bibfnamefont{P.~J.} \bibnamefont{Picone}},
  \bibinfo{author}{\bibfnamefont{M.~A.} \bibnamefont{Kastner}},
  \bibinfo{author}{\bibfnamefont{H.~P.} \bibnamefont{Jenssen}},
  \bibinfo{author}{\bibfnamefont{D.~R.} \bibnamefont{Gabbe}},
  \bibinfo{author}{\bibfnamefont{C.~Y.} \bibnamefont{Chen}},
  \bibinfo{author}{\bibfnamefont{R.~J.} \bibnamefont{Birgeneau}},
  \bibnamefont{and} \bibinfo{author}{\bibfnamefont{A.}~\bibnamefont{Aharony}},
  \bibinfo{journal}{Phys.\ Rev.\ B} \textbf{\bibinfo{volume}{38}},
  \bibinfo{pages}{905} (\bibinfo{year}{1988}).

\bibitem[{\citenamefont{H\"ucker et~al.}(2002)\citenamefont{H\"ucker, Klauss,
  and B\"uchner}}]{Huecker02b}
\bibinfo{author}{\bibfnamefont{M.}~\bibnamefont{H\"ucker}},
  \bibinfo{author}{\bibfnamefont{H.-H.} \bibnamefont{Klauss}},
  \bibnamefont{and}
  \bibinfo{author}{\bibfnamefont{B.}~\bibnamefont{B\"uchner}},
  \bibinfo{journal}{cond-mat} \textbf{\bibinfo{volume}{0210357}}
  (\bibinfo{year}{2002}).

\bibitem[{\citenamefont{Keimer et~al.}(1993)\citenamefont{Keimer, Birgeneau,
  Cassanho, Endoh, Greven, Kastner, and Shirane}}]{Keimer93}
\bibinfo{author}{\bibfnamefont{B.}~\bibnamefont{Keimer}},
  \bibinfo{author}{\bibfnamefont{R.~J.} \bibnamefont{Birgeneau}},
  \bibinfo{author}{\bibfnamefont{A.}~\bibnamefont{Cassanho}},
  \bibinfo{author}{\bibfnamefont{Y.}~\bibnamefont{Endoh}},
  \bibinfo{author}{\bibfnamefont{M.}~\bibnamefont{Greven}},
  \bibinfo{author}{\bibfnamefont{M.~A.} \bibnamefont{Kastner}},
  \bibnamefont{and} \bibinfo{author}{\bibfnamefont{G.}~\bibnamefont{Shirane}},
  \bibinfo{journal}{Z.\ Phys. B}
  \textbf{\bibinfo{volume}{91}}, \bibinfo{pages}{373} (\bibinfo{year}{1993}).

\bibitem[{\citenamefont{Thio and Aharony}(1994)}]{Thio94}
\bibinfo{author}{\bibfnamefont{T.}~\bibnamefont{Thio}} \bibnamefont{and}
  \bibinfo{author}{\bibfnamefont{A.}~\bibnamefont{Aharony}},
  \bibinfo{journal}{Phys.\ Rev.\ Lett.} \textbf{\bibinfo{volume}{73}},
  \bibinfo{pages}{894} (\bibinfo{year}{1994}).

\bibitem[{\citenamefont{Thio et~al.}(1990)\citenamefont{Thio, Chen, Freer,
  Gabbe, Jenssen, Kastner, Picone, Preyer, and Birgeneau}}]{Thio90}
\bibinfo{author}{\bibfnamefont{T.}~\bibnamefont{Thio}},
  \bibinfo{author}{\bibfnamefont{C.~Y.} \bibnamefont{Chen}},
  \bibinfo{author}{\bibfnamefont{B.~S.} \bibnamefont{Freer}},
  \bibinfo{author}{\bibfnamefont{D.~R.} \bibnamefont{Gabbe}},
  \bibinfo{author}{\bibfnamefont{H.~P.} \bibnamefont{Jenssen}},
  \bibinfo{author}{\bibfnamefont{M.~A.} \bibnamefont{Kastner}},
  \bibinfo{author}{\bibfnamefont{P.~J.} \bibnamefont{Picone}},
  \bibinfo{author}{\bibfnamefont{N.~W.} \bibnamefont{Preyer}},
  \bibnamefont{and} \bibinfo{author}{\bibfnamefont{R.~J.}
  \bibnamefont{Birgeneau}}, \bibinfo{journal}{Phys.\ Rev.\ B}
  \textbf{\bibinfo{volume}{41}}, \bibinfo{pages}{231} (\bibinfo{year}{1990}).

\bibitem[{\citenamefont{Vaknin et~al.}(1990)\citenamefont{Vaknin, Sinha,
  Stassis, Miller, and Johnston}}]{Vaknin90}
\bibinfo{author}{\bibfnamefont{D.}~\bibnamefont{Vaknin}},
  \bibinfo{author}{\bibfnamefont{S.~K.} \bibnamefont{Sinha}},
  \bibinfo{author}{\bibfnamefont{C.}~\bibnamefont{Stassis}},
  \bibinfo{author}{\bibfnamefont{L.~L.} \bibnamefont{Miller}},
  \bibnamefont{and} \bibinfo{author}{\bibfnamefont{D.~C.}
  \bibnamefont{Johnston}}, \bibinfo{journal}{Phys.\ Rev.\ B}
  \textbf{\bibinfo{volume}{41}}, \bibinfo{pages}{1926}
  (\bibinfo{year}{1990}).

\bibitem[{\citenamefont{Uchida et~al.}(1991)\citenamefont{Uchida, Ido, Takagi,
  Arima, Tokura, and Tajima}}]{Uchida91}
\bibinfo{author}{\bibfnamefont{S.}~\bibnamefont{Uchida}},
  \bibinfo{author}{\bibfnamefont{T.}~\bibnamefont{Ido}},
  \bibinfo{author}{\bibfnamefont{H.}~\bibnamefont{Takagi}},
  \bibinfo{author}{\bibfnamefont{T.}~\bibnamefont{Arima}},
  \bibinfo{author}{\bibfnamefont{Y.}~\bibnamefont{Tokura}}, \bibnamefont{and}
  \bibinfo{author}{\bibfnamefont{S.}~\bibnamefont{Tajima}},
  \bibinfo{journal}{Phys.\ Rev.\ B} \textbf{\bibinfo{volume}{43}},
  \bibinfo{pages}{7942} (\bibinfo{year}{1991}).

\bibitem[{\citenamefont{Shibata et~al.}(1993)\citenamefont{Shibata, Watanabe,
  Kinoshita, Matsuda, and Yamada}}]{Shibata93a}
\bibinfo{author}{\bibfnamefont{H.}~\bibnamefont{Shibata}},
  \bibinfo{author}{\bibfnamefont{T.}~\bibnamefont{Watanabe}},
  \bibinfo{author}{\bibfnamefont{K.}~\bibnamefont{Kinoshita}},
  \bibinfo{author}{\bibfnamefont{A.}~\bibnamefont{Matsuda}}, \bibnamefont{and}
  \bibinfo{author}{\bibfnamefont{T.}~\bibnamefont{Yamada}},
  \bibinfo{journal}{Phys.\ Rev.\ B} \textbf{\bibinfo{volume}{48}},
  \bibinfo{pages}{14027} (\bibinfo{year}{1993}).

\bibitem[{\citenamefont{Niedermayer et~al.}(1998)\citenamefont{Niedermayer,
  Bernhard, Blasius, Golnik, Moodenbaugh, and Budnick}}]{Niedermayer98}
\bibinfo{author}{\bibfnamefont{C.}~\bibnamefont{Niedermayer}},
  \bibinfo{author}{\bibfnamefont{C.}~\bibnamefont{Bernhard}},
  \bibinfo{author}{\bibfnamefont{T.}~\bibnamefont{Blasius}},
  \bibinfo{author}{\bibfnamefont{A.}~\bibnamefont{Golnik}},
  \bibinfo{author}{\bibfnamefont{A.}~\bibnamefont{Moodenbaugh}},
  \bibnamefont{and} \bibinfo{author}{\bibfnamefont{J.~I.}
  \bibnamefont{Budnick}}, \bibinfo{journal}{Phys.\ Rev.\ Lett.}
  \textbf{\bibinfo{volume}{80}}, \bibinfo{pages}{3843} (\bibinfo{year}{1998}).

\bibitem[{\citenamefont{Greven et~al.}(1995)\citenamefont{Greven, Birgeneau,
  Endoh, Kastner, Matsuda, and Shirane}}]{Greven95}
\bibinfo{author}{\bibfnamefont{M.}~\bibnamefont{Greven}},
  \bibinfo{author}{\bibfnamefont{R.~J.} \bibnamefont{Birgeneau}},
  \bibinfo{author}{\bibfnamefont{Y.}~\bibnamefont{Endoh}},
  \bibinfo{author}{\bibfnamefont{M.~A.} \bibnamefont{Kastner}},
  \bibinfo{author}{\bibfnamefont{M.}~\bibnamefont{Matsuda}}, \bibnamefont{and}
  \bibinfo{author}{\bibfnamefont{G.}~\bibnamefont{Shirane}},
  \bibinfo{journal}{Z.\ Phys. B}
  \textbf{\bibinfo{volume}{96}}, \bibinfo{pages}{465} (\bibinfo{year}{1995}).

\bibitem[{\citenamefont{Xue et~al.}(1988)\citenamefont{Xue, Grest, Cohen, and
  Sinha}}]{Xue88}
\bibinfo{author}{\bibfnamefont{W.}~\bibnamefont{Xue}},
  \bibinfo{author}{\bibfnamefont{G.~S.} \bibnamefont{Grest}},
  \bibinfo{author}{\bibfnamefont{M.~H.} \bibnamefont{Cohen}}, \bibnamefont{and}
  \bibinfo{author}{\bibfnamefont{S.~K.} \bibnamefont{Sinha}},
  \bibinfo{journal}{Phys.\ Rev.\ B} \textbf{\bibinfo{volume}{38}},
  \bibinfo{pages}{6868} (\bibinfo{year}{1988}).

\bibitem[{\citenamefont{Suh et~al.}(1995)\citenamefont{Suh, Borsa, Miller,
  Corti, Johnston, and Torgeson}}]{Suh95}
\bibinfo{author}{\bibfnamefont{B.~J.} \bibnamefont{Suh}},
  \bibinfo{author}{\bibfnamefont{F.}~\bibnamefont{Borsa}},
  \bibinfo{author}{\bibfnamefont{L.~L.} \bibnamefont{Miller}},
  \bibinfo{author}{\bibfnamefont{M.}~\bibnamefont{Corti}},
  \bibinfo{author}{\bibfnamefont{D.~C.} \bibnamefont{Johnston}},
  \bibnamefont{and} \bibinfo{author}{\bibfnamefont{D.~R.}
  \bibnamefont{Torgeson}}, \bibinfo{journal}{Phys.\ Rev.\ Lett.}
  \textbf{\bibinfo{volume}{75}}, \bibinfo{pages}{2212} (\bibinfo{year}{1995}).

\bibitem[{\citenamefont{Ansaldo et~al.}(1992)\citenamefont{Ansaldo,
  Niedermayer, Gl\"uckler, Stronach, Riseman, Noakes, Orbadors, Fuertes,
  Navarro, Gomez et~al.}}]{Ansaldo92a}
\bibinfo{author}{\bibfnamefont{E.~J.} \bibnamefont{Ansaldo}},
  \bibinfo{author}{\bibfnamefont{C.}~\bibnamefont{Niedermayer}},
  \bibinfo{author}{\bibfnamefont{H.}~\bibnamefont{Gl\"uckler}},
  \bibinfo{author}{\bibfnamefont{C.~E.} \bibnamefont{Stronach}},
  \bibinfo{author}{\bibfnamefont{T.~M.} \bibnamefont{Riseman}},
  \bibinfo{author}{\bibfnamefont{D.~R.} \bibnamefont{Noakes}},
  \bibinfo{author}{\bibfnamefont{X.}~\bibnamefont{Orbadors}},
  \bibinfo{author}{\bibfnamefont{A.}~\bibnamefont{Fuertes}},
  \bibinfo{author}{\bibfnamefont{J.~M.} \bibnamefont{Navarro}},
  \bibinfo{author}{\bibfnamefont{P.}~\bibnamefont{Gomez}},
  \bibnamefont{et~al.}, \bibinfo{journal}{Phys.\ Rev.\ B}
  \textbf{\bibinfo{volume}{46}}, \bibinfo{pages}{3084}
  (\bibinfo{year}{1992}).

\bibitem[{und()}]{undoped}
  \bibinfo{note}{The N\'eel temperature of pure \lcco\ with $\delta=0$
  is not known.}

\bibitem[{\citenamefont{Keimer et~al.}(1992)\citenamefont{Keimer, Belk,
  Birgeneau, Cassanho, Chen, Greven, Aharony, Endoh, Erwin, and
  Shirane}}]{Keimer92b}
\bibinfo{author}{\bibfnamefont{B.}~\bibnamefont{Keimer}},
  \bibinfo{author}{\bibfnamefont{N.}~\bibnamefont{Belk}},
  \bibinfo{author}{\bibfnamefont{R.~J.} \bibnamefont{Birgeneau}},
  \bibinfo{author}{\bibfnamefont{A.}~\bibnamefont{Cassanho}},
  \bibinfo{author}{\bibfnamefont{C.~Y.} \bibnamefont{Chen}},
  \bibinfo{author}{\bibfnamefont{M.}~\bibnamefont{Greven}},
  \bibinfo{author}{\bibfnamefont{M.~A. K.~A.} \bibnamefont{Aharony}},
  \bibinfo{author}{\bibfnamefont{Y.}~\bibnamefont{Endoh}},
  \bibinfo{author}{\bibfnamefont{R.~W.} \bibnamefont{Erwin}}, \bibnamefont{and}
  \bibinfo{author}{\bibfnamefont{G.}~\bibnamefont{Shirane}},
  \bibinfo{journal}{Phys.\ Rev.\ B} \textbf{\bibinfo{volume}{46}},
  \bibinfo{pages}{14034} (\bibinfo{year}{1992}).

\bibitem[{\citenamefont{Matsuda et~al.}(2000)\citenamefont{Matsuda, Lee,
  Greven, Kastner, Birgeneau, Yamada, Endoh, B\"{o}ni, Lee, Wakimoto
  et~al.}}]{Matsuda00a}
\bibinfo{author}{\bibfnamefont{M.}~\bibnamefont{Matsuda}},
  \bibinfo{author}{\bibfnamefont{Y.~S.} \bibnamefont{Lee}},
  \bibinfo{author}{\bibfnamefont{M.}~\bibnamefont{Greven}},
  \bibinfo{author}{\bibfnamefont{M.~A.} \bibnamefont{Kastner}},
  \bibinfo{author}{\bibfnamefont{R.~J.} \bibnamefont{Birgeneau}},
  \bibinfo{author}{\bibfnamefont{K.}~\bibnamefont{Yamada}},
  \bibinfo{author}{\bibfnamefont{Y.}~\bibnamefont{Endoh}},
  \bibinfo{author}{\bibfnamefont{P.}~\bibnamefont{B\"{o}ni}},
  \bibinfo{author}{\bibfnamefont{S.-H.} \bibnamefont{Lee}},
  \bibinfo{author}{\bibfnamefont{S.}~\bibnamefont{Wakimoto}},
  \bibnamefont{et~al.}, \bibinfo{journal}{Phys.\ Rev.\ B}
  \textbf{\bibinfo{volume}{61}}, \bibinfo{pages}{4326}
  (\bibinfo{year}{2000}).

\bibitem[{\citenamefont{Hiraka et~al.}(2001)\citenamefont{Hiraka, Endoh,
  Fujita, Lee, Kulda, Ivanov, and Birgeneau}}]{Hiraka01a}
\bibinfo{author}{\bibfnamefont{H.}~\bibnamefont{Hiraka}},
  \bibinfo{author}{\bibfnamefont{Y.}~\bibnamefont{Endoh}},
  \bibinfo{author}{\bibfnamefont{M.}~\bibnamefont{Fujita}},
  \bibinfo{author}{\bibfnamefont{Y.~S.} \bibnamefont{Lee}},
  \bibinfo{author}{\bibfnamefont{J.}~\bibnamefont{Kulda}},
  \bibinfo{author}{\bibfnamefont{A.}~\bibnamefont{Ivanov}}, \bibnamefont{and}
  \bibinfo{author}{\bibfnamefont{R.~J.} \bibnamefont{Birgeneau}},
  \bibinfo{journal}{J.\ Phys.\ Soc.\ Japan} \textbf{\bibinfo{volume}{70}},
  \bibinfo{pages}{853} (\bibinfo{year}{2001}).

\bibitem[{\citenamefont{Takagi et~al.}(1992)\citenamefont{Takagi, Batlogg, Kao,
  Kwo, Cava, Krajewski, and {Peck Jr.}}}]{Takagi92b}
\bibinfo{author}{\bibfnamefont{H.}~\bibnamefont{Takagi}},
  \bibinfo{author}{\bibfnamefont{B.}~\bibnamefont{Batlogg}},
  \bibinfo{author}{\bibfnamefont{H.~L.} \bibnamefont{Kao}},
  \bibinfo{author}{\bibfnamefont{J.}~\bibnamefont{Kwo}},
  \bibinfo{author}{\bibfnamefont{R.~J.} \bibnamefont{Cava}},
  \bibinfo{author}{\bibfnamefont{J.~J.} \bibnamefont{Krajewski}},
  \bibnamefont{and} \bibinfo{author}{\bibfnamefont{W.~F.} \bibnamefont{{Peck
  Jr.}}}, \bibinfo{journal}{Phys.\ Rev.\ Lett.} \textbf{\bibinfo{volume}{69}},
  \bibinfo{pages}{2975} (\bibinfo{year}{1992}).

\bibitem[{\citenamefont{Chou et~al.}(1993)\citenamefont{Chou, Borsa, Cho,
  Johnston, Lascialfari, Torgeson, and Ziolo}}]{Chou93}
\bibinfo{author}{\bibfnamefont{F.~C.} \bibnamefont{Chou}},
  \bibinfo{author}{\bibfnamefont{F.}~\bibnamefont{Borsa}},
  \bibinfo{author}{\bibfnamefont{J.~H.} \bibnamefont{Cho}},
  \bibinfo{author}{\bibfnamefont{D.~C.} \bibnamefont{Johnston}},
  \bibinfo{author}{\bibfnamefont{A.}~\bibnamefont{Lascialfari}},
  \bibinfo{author}{\bibfnamefont{D.~R.} \bibnamefont{Torgeson}},
  \bibnamefont{and} \bibinfo{author}{\bibfnamefont{J.}~\bibnamefont{Ziolo}},
  \bibinfo{journal}{Phys.\ Rev.\ Lett.} \textbf{\bibinfo{volume}{71}},
  \bibinfo{pages}{2323} (\bibinfo{year}{1993}).

\bibitem[{\citenamefont{Lai and Gooding}(1998)}]{Gooding98a}
\bibinfo{author}{\bibfnamefont{E.}~\bibnamefont{Lai}} \bibnamefont{and}
  \bibinfo{author}{\bibfnamefont{R.~J.} \bibnamefont{Gooding}},
  \bibinfo{journal}{Phys.\ Rev.\ B} \textbf{\bibinfo{volume}{57}},
  \bibinfo{pages}{1498} (\bibinfo{year}{1998}).

\bibitem[{\citenamefont{Kinoshita et~al.}(1992)\citenamefont{Kinoshita, Izumi,
  Yamada, and Asano}}]{Kinoshita92a}
\bibinfo{author}{\bibfnamefont{K.}~\bibnamefont{Kinoshita}},
  \bibinfo{author}{\bibfnamefont{F.}~\bibnamefont{Izumi}},
  \bibinfo{author}{\bibfnamefont{T.}~\bibnamefont{Yamada}}, \bibnamefont{and}
  \bibinfo{author}{\bibfnamefont{H.}~\bibnamefont{Asano}},
  \bibinfo{journal}{Phys.\ Rev.\ B} \textbf{\bibinfo{volume}{45}},
  \bibinfo{pages}{5558} (\bibinfo{year}{1992}).

\bibitem[{\citenamefont{Wakimoto et~al.}(1999)\citenamefont{Wakimoto, Endoh,
  Hirota, Ueki, Yamada, Birgeneau, Kastner, Lee, Gehring, and
  Lee}}]{Wakimoto99a}
\bibinfo{author}{\bibfnamefont{S.}~\bibnamefont{Wakimoto}},
  \bibinfo{author}{\bibfnamefont{G.~S.~Y.} \bibnamefont{Endoh}},
  \bibinfo{author}{\bibfnamefont{K.}~\bibnamefont{Hirota}},
  \bibinfo{author}{\bibfnamefont{S.}~\bibnamefont{Ueki}},
  \bibinfo{author}{\bibfnamefont{K.}~\bibnamefont{Yamada}},
  \bibinfo{author}{\bibfnamefont{R.~J.} \bibnamefont{Birgeneau}},
  \bibinfo{author}{\bibfnamefont{M.~A.} \bibnamefont{Kastner}},
  \bibinfo{author}{\bibfnamefont{Y.~S.} \bibnamefont{Lee}},
  \bibinfo{author}{\bibfnamefont{P.~M.} \bibnamefont{Gehring}},
  \bibnamefont{and} \bibinfo{author}{\bibfnamefont{S.~H.} \bibnamefont{Lee}},
  \bibinfo{journal}{Phys.\ Rev.\ B} \textbf{\bibinfo{volume}{60}},
  \bibinfo{pages}{R769} (\bibinfo{year}{1999}).

\end{thebibliography}
\end{document}